\documentclass[showpacs,twocolumn,superscriptaddress]{revtex4-2}
\usepackage{graphicx}
\usepackage{amssymb}
\usepackage{amsmath}
\usepackage{array}
\usepackage{color}
\usepackage{hyperref} 
\hypersetup{colorlinks=true,linkcolor=blue,citecolor=cyan}

\begin{document}

\title{Electric Penrose process and the accretion disk around a 4D charged Einstein-Gauss-Bonnet black hole}

\author{Mirzabek Alloqulov}
\email{malloqulov@gmail.com}
\affiliation{Institute of Fundamental and Applied Research, National Research University TIIAME, Kori Niyoziy 39, Tashkent 100000, Uzbekistan} 
\affiliation{University of Tashkent for Applied Sciences, Str. Gavhar 1, Tashkent 100149, Uzbekistan}

\author{Sanjar Shaymatov}
\email{sanjar@astrin.uz}
\affiliation{Institute for Theoretical Physics and Cosmology,
Zhejiang University of Technology, Hangzhou 310023, China}
\affiliation{Institute of Fundamental and Applied Research, National Research University TIIAME, Kori Niyoziy 39, Tashkent 100000, Uzbekistan}
\affiliation{University of Tashkent for Applied Sciences, Str. Gavhar 1, Tashkent 100149, Uzbekistan}

%
\date{\today}
\begin{abstract}

In this paper, we aim to examine the electric Penrose process (PP) around a charged black hole in 4D Einstein–Gauss–Bonnet (EGB) gravity and bring out the effect of the Gauss–Bonnet (GB) coupling parameter $\alpha$ and black hole charge on the efficiency of energy extraction from the black hole. This research is motivated by the fact that electrostatic interactions significantly influence the behavior of charged particles in the vicinity of a charged static black hole. Under this interaction, decaying charged particles can have negative energies, causing energy to be released from black holes with no ergosphere. We show that the GB coupling parameter has a significant impact on the energy efficiency of the electric PP, but the efficiency can be strongly enhanced by the black hole charge due to the Coulomb force. Finally, we consider the accretion disk around the black hole and investigate in detail its radiation properties, such as the electromagnetic radiation flux, the temperature, and the differential luminosity. We show that the GB coupling parameter can have a significant impact on the radiation parameters, causing them to increase in the accretion disk in the vicinity of the black hole. Interestingly, it is found that the 4D EGB charged black hole is more efficient and favorable for the accretion disk radiation compared to a charged black hole in Einstein gravity.   

\end{abstract}

\maketitle
\footnotesize

\section{Introduction}

In general relativity (GR), black holes have been known as a generic result of Einstein's gravity, as their geometric properties are described by simple mathematical equations.  Among other properties the occurrence of singularity makes black holes very exciting and fascinating objects, even though this marks the limits of Einstein's theory.  However, exploring black hole's unknown properties and the limits of the theory's applicability has been extremely important. In this context general relativity is known as an incomplete theory and thus for its validity and applicability higher-order theories have been regarded as possible extensions of general relativity~\cite{Dadhich12c}. We know that Einstein's gravity is constructed from linear order Riemann curvature, while Gauss-Bonnet (GB) gravity continues to the quadratic order with higher order invariants, thus referring to the Lovelock theory as a generalization of Einstein's theory~\cite{Lovelock1971}. Note that GB gravity with the quadratic order gives a contribution to the gravitational dynamics only in $D>4$, and hence this is known as the Lovelock theory. However, it was recently suggested that a 4-dimensional Einstein-Gauss-Bonnet ($4D$ EGB) theory could exist and that Lovelock's theorem could be bypassed by a suitable redefinition of the GB coupling constant~\cite{Glavan20prl}. The new $4D$ EGB is presently under scrutiny on the basis of two main arguments. One involving the ill-posedness of the action for the theory~\cite{Gurses20egb,Mahapatra20egb} and the other regarding the validity of the rescaling of the GB constant, which may be possible only for systems with certain symmetries~\cite{Hennigar20egb}. Both objections, if valid, may invalidate the $4D$ EGB theory as an alternative to Einstein's theory. However, the fact remains that solutions with high symmetries exist, they have a clear physical interpretation that mirrors the corresponding solutions in GR, and may be regarded as coming from an effective prescription for the lower dimensional limit of GB gravity. For this reason, the investigation of the properties of such solutions has attracted great interest (see, e.g. ~\cite{Liu20egb,Guo20egb,Wei20egb,Kumar20egb,Konoplya20egb,Churilova20egb,Malafarina20egb,Aragon20egb,Mansoori20egb,Rayimbaev2020egb,Ge20egb,Chakraborty20egb,Odintsov20plb,Lin20egb,Aoki20egb,Shaymatov20egb,Islam20egb,Singh20-egb}). There have also been several investigations on these lines ~\cite{Zhang20egb} addressing the dynamics of spinning particle motion, the impact of the GB term  on the superradiance process~\cite{Zhang20aegb}, the plasma effect on weak gravitational lensing~\cite{Atamurotov21JCAP,Narzilloev21BTZ}, the question of destroying the black hole horizon in strong and weak forms~\cite{Mishra20egb,Yang20b,Ayyesha22PDU}, and the Bondi-Hoyle accretion process around $4D$ EGB black hole~\cite{Donmez2021egb}. It is worth noting that the $4D$ EGB theory was also extended to obtain black hole solutions with electric charge and rotation \cite{Fernandes20plb,Kumar20egb}. Later, it was also extended to the 3D BTZ black hole solution in EGB theory~\cite{Hennigar20PLB}. Additionally, interesting aspects that pertain to the GB black hole in higher dimensions (i.e., $D>4$ ) have also been investigated in Refs.~\cite{Shaymatov22JCAP,Dadhich22a,Wu21egb}. 

The Penrose process (PP) \cite{Penrose:1969pc} is a mechanism proposed to extract the rotational energy of rapidly rotating black holes, usually referred to as a potential explanation for highly energetic astrophysical phenomena. It utilizes the ergosphere, which appears in the region between the horizon and the static limit bounded from the outer surface. In this process, a falling particle is divided into two parts, with one part falling into the black hole and the other escaping to infinity with more energy than the incident particle. This allows the energy of the escaping particle to be extracted from the black hole, leading to a slowdown in the black hole's rotation. The PP has since been applied in various contexts \cite{Abdujabbarov11,Toshmatov:2014qja,Okabayashi20}. It is to be emphasized that the PP has also been extended to higher dimensional rotating black holes \cite{Prabhu10,Nozawa05,Shaymatov:2024MPP1} and Buchdahl stars \cite{Shaymatov:2024MPP2}. Bardeen et al. \cite{Bardeen72} and Wald \cite{Wald74ApJ} demonstrated that extracting more energy from the black hole would be difficult unless the incident particle is relativistic. Subsequently, the PP was reformulated as a new mechanism, known as the magnetic Penrose process \cite{Bhat85,Parthasarathy86}. In this mechanism, the influence of the magnetic field on an escaping particle enables it to surpass the constraint velocity and become relativistic, significantly improving the efficiency of energy extraction from a rotating Kerr black hole. This mechanism has since been extended to various scenarios  \cite{Wagh89,Alic12ApJ,Moesta12ApJ,Dadhich18mnras,Tursunov:2019oiq,Tursunov20ApJ,Shaymatov22b}) addressing the impacts of a purely magnetic field on the energy extraction process. Hence, it has been accepted that the energy extraction process can only occur inside the ergoregion of rotating black holes acting as an engine for the higher energetic astrophysical processes mentioned above. Despite this fact, the PP can be applied even in non-rotating black holes, referred to as the electric Penrose process, which can also serve as a high-energy emission event \cite{Bhat85,Denardo73PLB,Tursunov21EPP,Stuchlik:2021,ALLOQULOV2023302,Baez2024epp}. Therefore, it is also important to examine the electric PP to gain a deeper explanation of high-energy astrophysical phenomena.

In this paper, we study a 4D charged EGB black hole and explore the efficiency of energy extraction using the electric PP. This is our main focus for investigation, with further analysis of the accretion disk properties around the black hole. Additionally, we examine test particle dynamics and the innermost stable circular orbit (ISCO) around the black hole, along with the accretion disk's radiative energy efficiency in 4D EGB theory. 

It is commonly accepted that the radiation of the accretion disk and recent observations pertaining to particle outflows ~\cite{Fender04mnrs,Auchettl17ApJ,IceCube17b} have been tested to examine the astrophysical aspects of black holes. Therefore, to gain a deeper understanding of gravity and to test it in the strong field regime it is worth analyzing the rich observational phenomenology of electromagnetic radiation with expected thermal spectra through a thin accretion disk \cite{Abramowicz13}. In this sense, it becomes important to explore the electromagnetic radiation by a thin accretion disk in the close vicinity of black holes. In the current paper, we investigate the radiative properties of the aforementioned thin accretion disk using the geometrically thin and optically thick Novikov-Thorn disk model in the strong field regime of 4D charged EGB black hole. We also provide a precise comparison with the standard Reissner-Nordström black hole in Einstein gravity. 

The paper is organized as follows. We briefly review a charged black hole in 4D EGB theory in Sec.~\ref{Formalism}. We consider the electric Penrose process with a test particle dynamics in Sec.~\ref{Sec:EPP}. We study the geometrically thin Novikov-Thorne model for the accretion disk around the black hole in 4D EGB theory in and further discuss the flux of the radiant energy over the accretion disk, accretions disk's radiative efficiency, temperature profile and differential luminosity in Sec.~\ref{Sec:Model}. We end up with conclusion in Sec.~\ref{Sec:Summary}. Throughout this work, we use the signature $(–, +, +, +)$ for the spacetime metric and system of units $G=c=1$.

\section{Spacetime geometry in the 4D EGB gravity }\label{Formalism}
The action for EGB gravity can be defined by the following  action~\cite{Glavan20prl,Zhang20aegb}
\begin{align}
{\cal S}=\frac{1}{16\pi}\int d^Dx\sqrt{-g}\left(R+\frac{\alpha}{D-4}{\cal G}^2-F_{\alpha\beta}F^{\alpha\beta}\right)\, ,
\end{align}
where $\alpha$ refers to Gauss-Bonnet (GB) coupling parameter, while ${\cal G}$ to the GB invariant defined as
\begin{align}
{\cal G}^2=R^2-4R^{\alpha\beta}R_{\alpha\beta}+R^{\alpha\beta\mu\nu}R_{\alpha\beta\mu\nu}\, , 
\end{align}
where $R$ is the Ricci scalar and $R_{\alpha\beta}$ and $R_{\alpha\beta\mu\nu}$ the Ricci and Riemann tensors. Based on the action underlined above, the line element describing the static black hole spacetime in 4D EGB theory is given by ~\cite{Zhang20aegb,Glavan20prl}
\begin{align}\label{metric}
ds^2=-f(r)\,dt^2+\frac{dr^2}{f(r)}+r^2\left(d\theta^2+\sin^2\theta d\phi^2\right)\ ,
\end{align}
with metric function 
\begin{align}\label{lapse}\nonumber
f(r)&=1+\frac{r^2}{2\alpha}\left(1-\sqrt{1+\frac{4\alpha}{r^2}\left(\frac{2M}{r}-\frac{Q^2}{r^2}\right)}\right)\\&=\frac{2(r^2-2Mr+Q^2+\alpha)}{r^2+2\alpha+\sqrt{r^4+4\alpha\left(2Mr-Q^2\right)}}\, ,
\end{align}
where $M$ is the total mass and $Q$ the black hole electric charge. The corresponding vector potential is written as follows:
\begin{align}
A_\alpha=\left(-\frac{Q_e}{r},0,0,Q_m\cos\theta\right)\, .    
\end{align}

The parameter range of the coupling constant $\alpha$ can be found by imposing the minimum condition of $f(r)$, i.e., it is given as $0\leq\alpha/M^2\leq \sqrt{1-\alpha}$. It is obvious that one can recover the Reissner-Nordstrom solution in the limit of $\alpha\to 0$, so it is given by 
\begin{align}
\lim_{\alpha\to 0}f=1-\frac{2M}{r}+\frac{Q^2}{r^2}\, .   
\end{align}
In the case of $\alpha\ll1$, it does however take the following form 
\begin{align}
f\simeq 1-\frac{2M}{r}+\frac{Q^2}{r^2}+\frac{4\alpha M}{r^3}+{\cal O}(\alpha^2)\ .  
\end{align}
Interestingly, albeit the complicated form of $f(r)$, the horizon radius takes a simple form as $r_{\pm}=M+\sqrt{M^2-Q^2-\alpha}$, similarly to the horizon cases of RN or Kerr black hole spacetime in Einstein gravity. It must be noted that, in the equatorial plane (i.e., $\theta=\pi/2$), the Faraday tensor, $F_{\mu \nu}=A_{\nu,\mu}-A_{\mu,\nu}$, can be defined by only one independent nonzero component as 
\begin{equation}
    F_{tr}=-F_{rt}=-\dfrac{Q}{r^2}\, . 
\end{equation}


\section{Electric Penrose process and the extracted energy }\label{Sec:EPP}

Here we consider the electric PP to examine energy extraction from a charged black hole in 4D EGB gravity. It must be emphasized that energy extraction from rapidly rotating black holes remains one of the most important issues in astrophysics. PP was proposed as a key explanation for higher energetic astrophysical phenomena, such as the active galactic nuclei with luminosity of approximately $\sim 10^{45}\rm erg\cdot s^{-1}$. For the PP to be valid, there must exist an ergosphere, in which an incident particle splits into two fragments, with one falling into the black hole and the other escaping to infinity with more energy. However, this is not true for a non-rotating black hole. One can consider electrostatic interaction in the so-called generalized ergosphere around a static charged black hole, where it is assumed that decaying charged particles can have negative energies under the electrostatic interaction. This is how the electric PP can become efficient, releasing energy from charged black holes with no ergosphere.

\begin{figure*}
    \centering
    \includegraphics[scale=0.5]{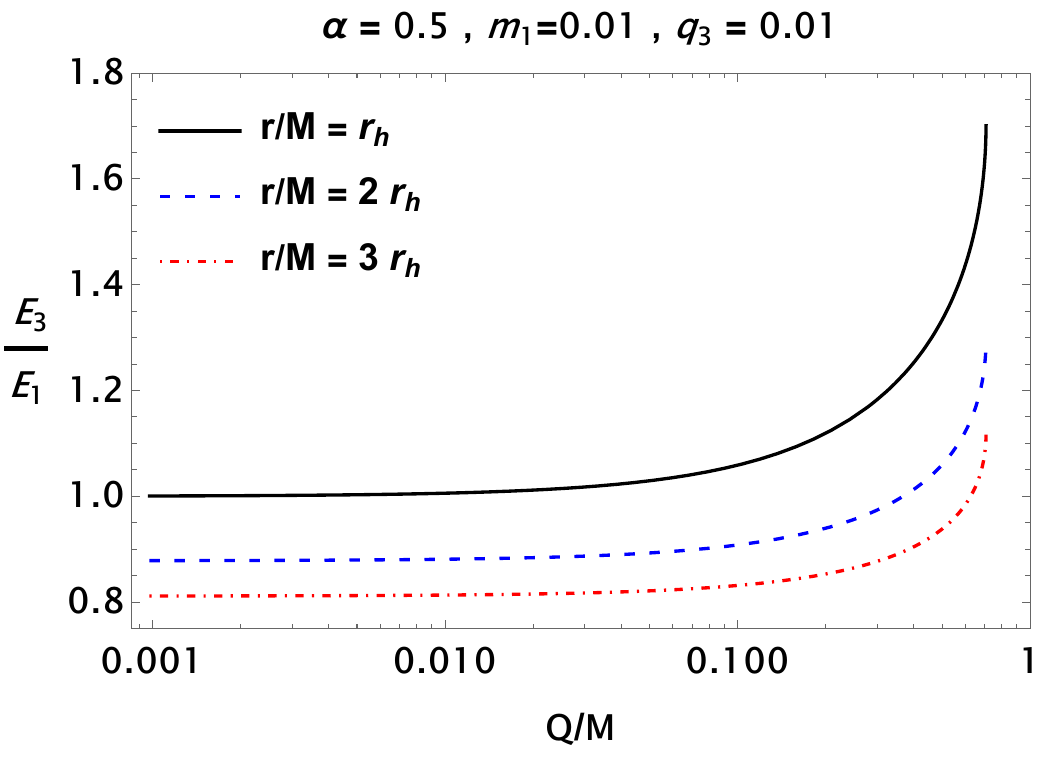}
    \includegraphics[scale=0.5]{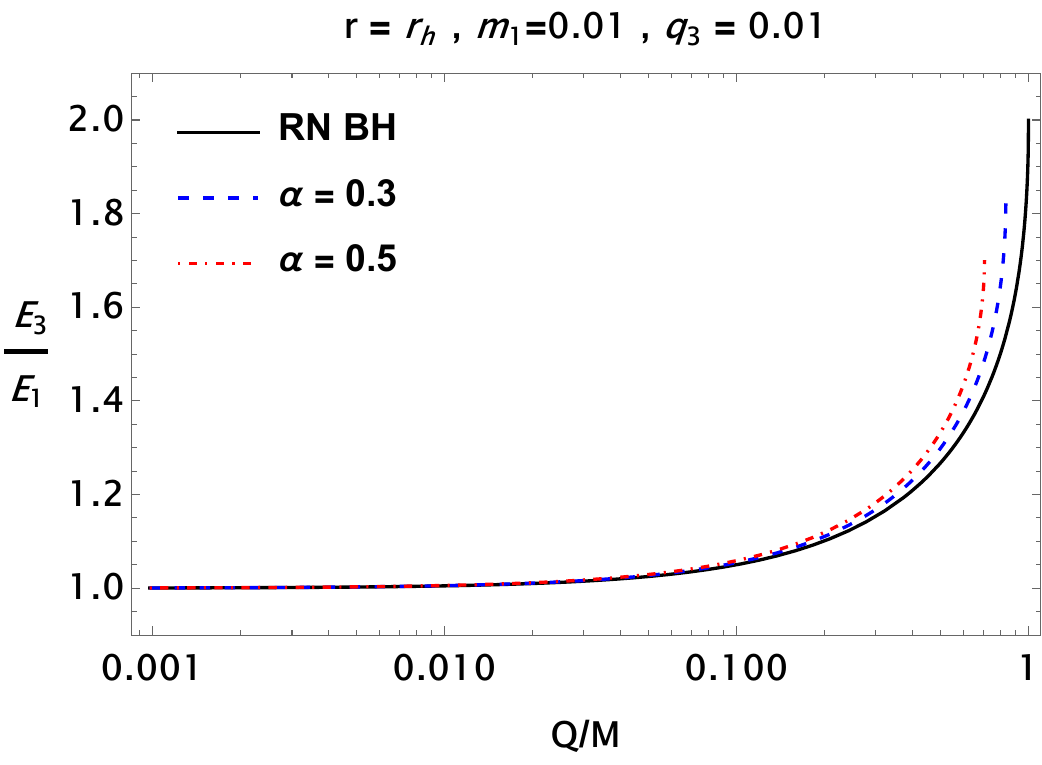}
    \caption{\label{fig:E3E1} The energy ratio (i.e., the energy efficiency of the electric PP) between ionized and neutral particles as a function of black hole charge $Q$.  Left panel: the energy efficiency is plotted for different values of the ionization points $r/M$. Right panel: the energy efficiency is plotted for different values of the GB coupling parameter $\alpha$. Here we set $m_1=0.01$ and $q_3=0.01$ for particle parameters.}
    
\end{figure*}

We assume that the splitting point of an incident particle takes place in the equatorial plane, with the four-velocity $u^{\alpha}=u^t(1,v,0,\Omega)$, where $v=dr/dt$ is the radial velocity of the particle and $\Omega=d\phi/dt$ the angular velocity. We can write the following equation for the angular velocity of the splitting particles using the normalization condition $u^{\alpha}u_{\alpha}=-k$ with $k=0$ for massless and $k=1$ for massive particle:
\begin{equation}\label{Eq:p1}
    (u^t)^2\left[\frac{v^2}{f(r)}-f(r)+\Omega^2 r^2\right]=-k\ .
\end{equation}
Based on the above equation, for a distant observer at infinity, the angular velocity of the splitting particles, $\Omega=d\phi/dt$, is defined by 
\begin{equation}\label{AngVel}
\Omega=\pm\frac{1}{u_t{r}}\sqrt{(u_t)^2(f(r)-f^{-1}(r)v^2)-k{f^2}(r)}\, .
\end{equation}
The allowed range of $\Omega$ is then given by  
\begin{equation}
\Omega_{-}\leq\Omega\leq\Omega_+\qquad \mbox{where} \qquad
\Omega_\pm=\pm\frac{\sqrt{f(r)}}{r}\, ,
\end{equation}
with the Keplerian orbits.
Further we consider the conservation laws for the incident particle, $m_1$, which splits into two fragments, $m_2$ and $m_3$, especially very close the black hole horizon. The conservation laws after splitting of the incident particle are then written as follows: 
\begin{equation}\label{EnAMq}
E_1 = E_2 + E_3\, ,\, \, \, 
L_1 = L_2 + L_3\,  \mbox{~~and~~}
q_1 = q_2 + q_3\, ,
\end{equation}
and 
\begin{equation}\label{Momentum}
m_1\dot{r_1} = m_2\dot{r_2} + m_3\dot{r_3},\qquad
m_1 \geq m_2 + m_3\, ,    
\end{equation}
where dot denotes a derivative with respect to the proper time, $\tau$. Following to the conservation laws, we write the momentum as \cite{Bhat85}
\begin{equation}\label{mphi}
m_1{u_1^\phi} = m_2{u_2^\phi} + m_3{u_3^\phi}\, ,
\end{equation}
with the four-velocity $u^\phi=\Omega{u^t}=\Omega{e}/f(r)$, where $i = 1,2,3$ depicts the number of particles in the process. Eq.~(\ref{mphi}) then yields 
\begin{equation}\label{me}
\Omega_1m_1e_1=\Omega_2m_2e_2+\Omega_3m_3e_3\, .
\end{equation}
Eq.~(\ref{me}) solves to give the analytic form of $E_3$ as
\begin{equation}\label{Energy3}
E_3=\frac{\Omega_1-\Omega_2}{\Omega_3-\Omega_2}(E_1 + q_1A_t)-q_3A_t,
\end{equation}
with the $i$th particle's angular velocity $\Omega_i=d\phi_i/dt$.

For the further analysis we shall for simplicity consider neutral incident particle (i.e., $q_1 = 0$) in order to maximize the ionized particle's energy. We also assume $E_1/m_1=1$. With this in view, for the incident particle $m_1$, the angular velocity given in Eq.~(\ref{AngVel}) can be rewritten as follows:
\begin{eqnarray}
&&\Omega_1^2=\frac{f(r)\left[1-f(r)\right]}{r^2}, \\
&& \Omega_2=\Omega_-\, ,\\
&& \Omega_3=\Omega_+\, .
\end{eqnarray}
with $q_2=-q_3$.
The particle $m_3$ can be considered an ionized particle.  This particle can reach its maximum energy, depending on the highest value of this ratio $(\Omega_1 - \Omega_2) / (\Omega_3 - \Omega_2)$ with the corresponding values of angular momenta $\Omega_{i}$. So one can then find   
\begin{equation} \label{AMofFragm}
\frac{\Omega_1-\Omega_2}{\Omega_3-\Omega_2}=\frac{1}{2}\Big[1+\sqrt{1-f(r_{ion})}\Big]\, , 
\end{equation}
where $r_{ion}$ refers to the ionization radius.
In the limit of $M=1$ and $Q=\alpha=0$, it reduces the Schwarzschild black hole case with the following form~\cite{Tursunov:2021,Kurbonov2023}
\begin{equation}
    \frac{\Omega_1-\Omega_2}{\Omega_3-\Omega_2}=\frac{1}{2}+\frac{1}{\sqrt{2 r_{ion}}}\, .
\end{equation}
Taking all into account, the ionized particle's energy take the following form 
\begin{equation}\label{energy1}
E_3=\frac{1}{2}\Big[1+\sqrt{1-f(r_{ion})}\Big] (E_1 + q_1 A_t) - q_3 A_t\, .
\end{equation}
As underlined above, we assume $q_1=0$ and $q_2=-q_3$. Within this assumption, the ionized particle's energy takes the following form at the splitting point 
\begin{equation}\label{energy2}
    E_3 = \frac{1}{2}\Big[1+\sqrt{1-f(r_{ion})}\Big] E_1 - q_3 A_t\, ,
\end{equation}
which, on the other hand, is rewritten as follows: 
\begin{equation}\label{E3E1}
    \frac{E_3}{E_1} = \frac{1}{2}\Big[1+\sqrt{1-f(r_{ion})}\Big]  - \frac{ q_3 A_t}{E_1}\, .
\end{equation}
The above equation represents the energy efficiency of the electric PP. We do further analyze the efficiency of energy extraction from a charged black hole in 4D EGB gravity. It is clear from the above equation that the time component of the electromagnetic four potential is properly proportional to the black hole charge $Q$. Therefore, the energy of the ionized particle becomes maximum when considering the sign for $q_3$ and $Q$ as expected, resulting in the charged particle getting accelerated due to the Coulomb repulsion force acting between the black hole charge and particle charge.

The key point to note here is that the ionized particle gets accelerated only if the right-hand side of Eq.~(\ref{E3E1}) is greater than unity. Since it is complicated to solve analytically we explore it numerically and provide illustrative plots. We then analyze the energy efficiency as the function of black hole charge for various possible cases. In Fig. \ref{fig:E3E1}, we demonstrate the energy efficiency released from the charged black hole in 4D EGB gravity due to the incident particle splitting into fragments in the generalized ergosphere. To be more precise, the left panel depicts the impact of black hole charge on the energy efficiency at different splitting points for fixed GB coupling parameter, while the right panel shows the same behavior at the fixed position for various combinations of GB coupling parameter. As can be seen from the left panel of Fig.~\ref{fig:E3E1}, the shape
of the energy efficiency (ratio) shifts up to higher efficiency when the splitting point occurs near the black hole horizon. Additionally, the energy efficiency increases with increasing black hole charge due to the increase in the Coulomb force. However, it decreases as the GB coupling parameter increases, compared to the RN black hole in Einstein gravity.      
\begin{figure*}
\centering
  \includegraphics[scale=0.5]{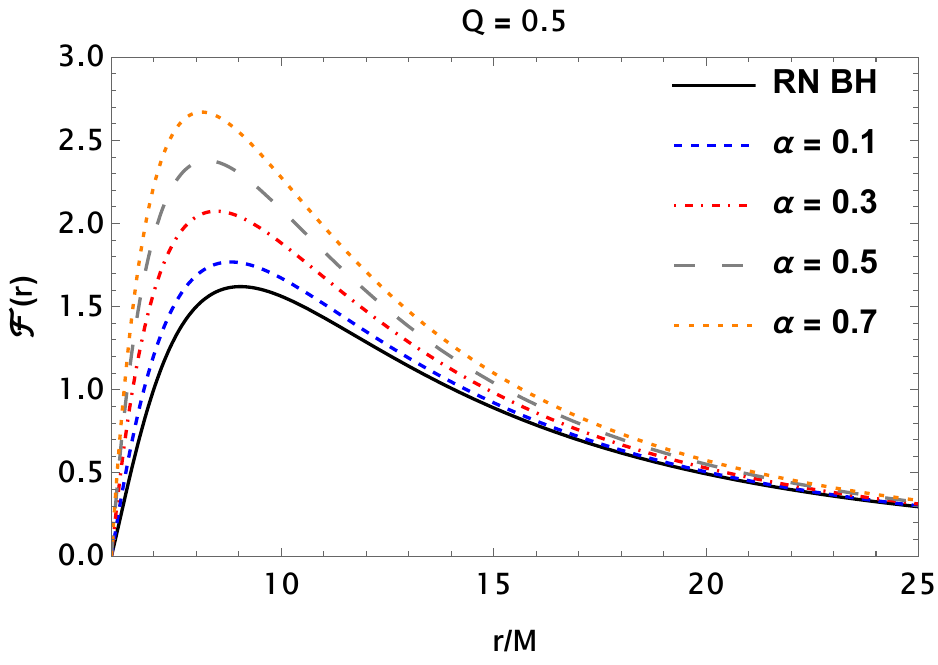}
 \includegraphics[scale=0.5]{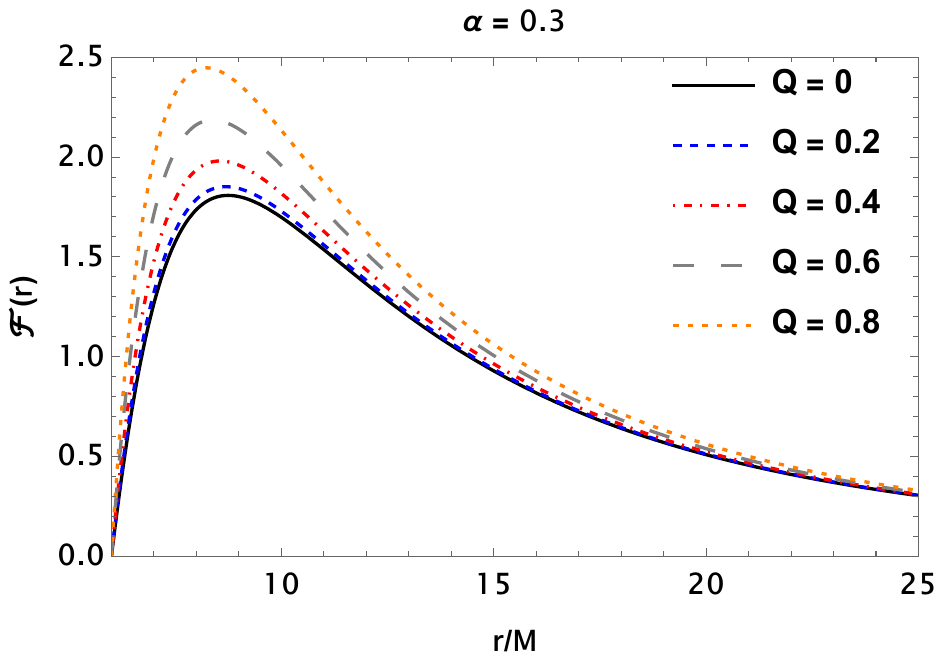}
\caption{\label{fig:flux} The plot shows the radial dependence of the electromagnetic radiation flux of the accretion disk for different possible cases of $\alpha$ and $Q$. Here, we note that for analysis, we consider the flux $\mathcal{F}$ of the accretion disk to be on the order of $10^{-5}$.}
\end{figure*}  
\begin{figure}
    \centering
    \includegraphics[scale=0.55]{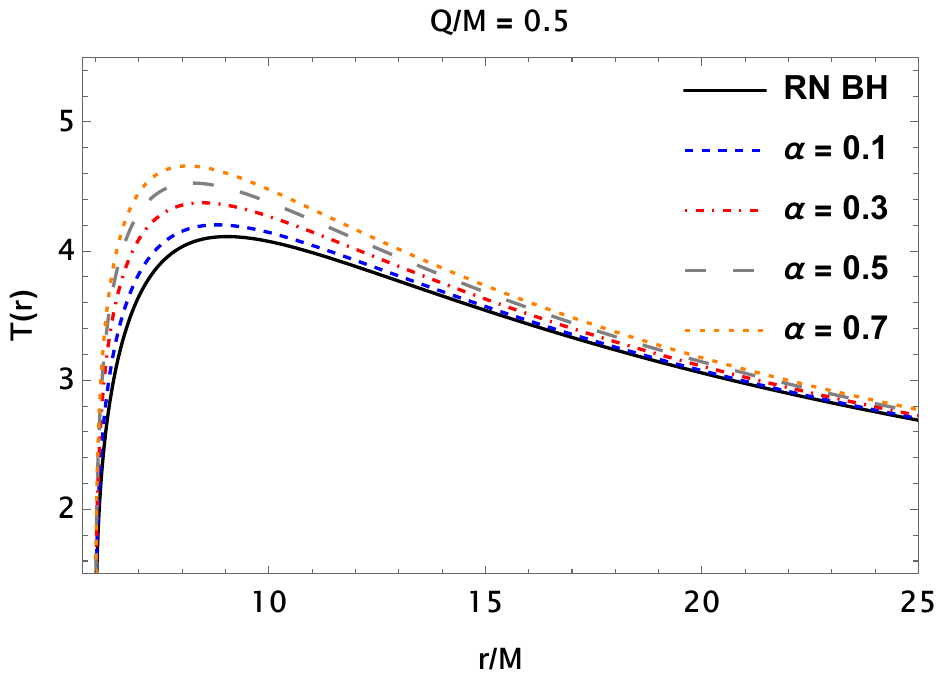}
   
        \caption{The plot shows the radial dependence of the accretion disk temperature for different possible case of the GB coupling parameter $\alpha$ for fixed $Q$. }
    \label{temperature1}
\end{figure}
\begin{figure*}
    \centering
    \includegraphics[scale=0.55]{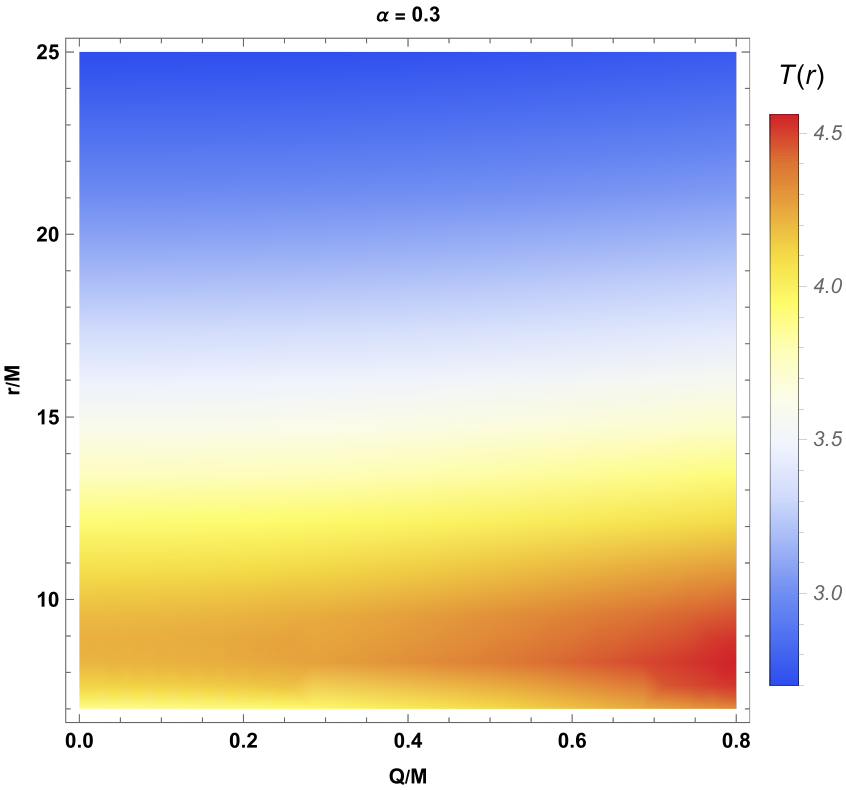}
    \includegraphics[scale=0.55]{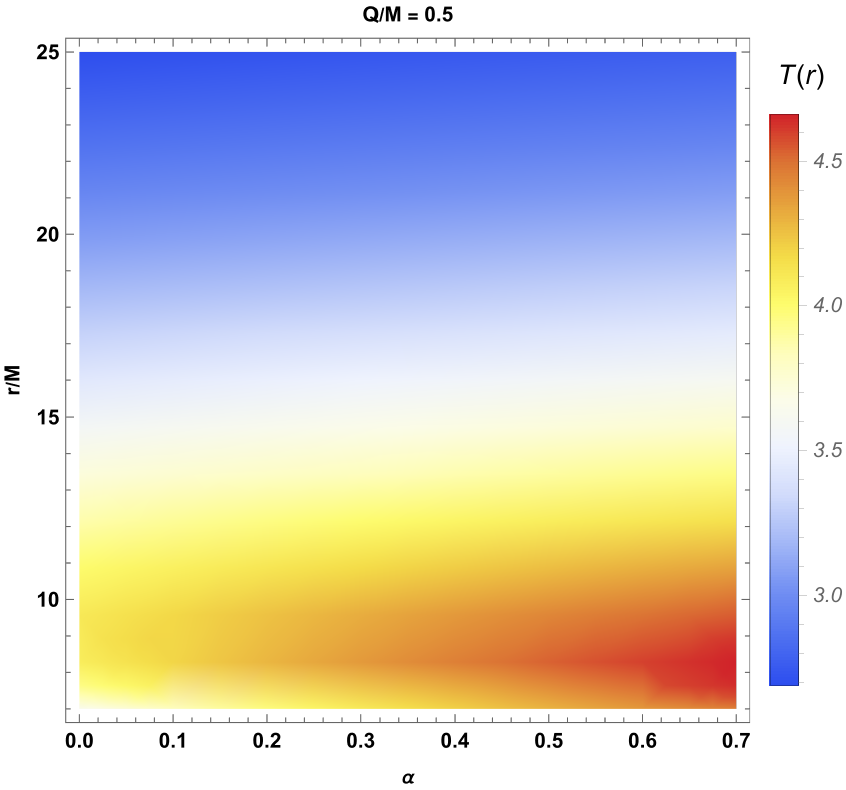}
    \includegraphics[scale=0.55]{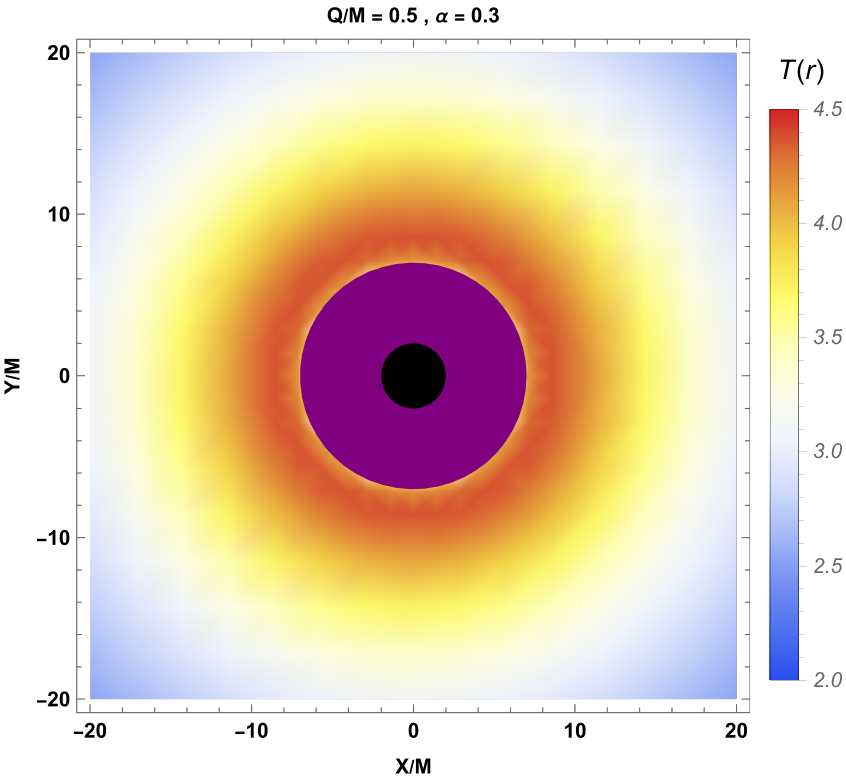}
    \caption{Plot shows the temperature profile of the accretion disk. The top left and right panels show the temperature profile in the parameter space of ($r,Q$) and ($r,\alpha$). The bottom panel shows the density plot of the disk temperature at the equatorial $X-Y$ plane, where $X$ and $Y$ refer to the Cartesian coordinates.}
    \label{temperature2}
\end{figure*}

\begin{figure}[t]
    \centering
    \includegraphics[scale=0.55]{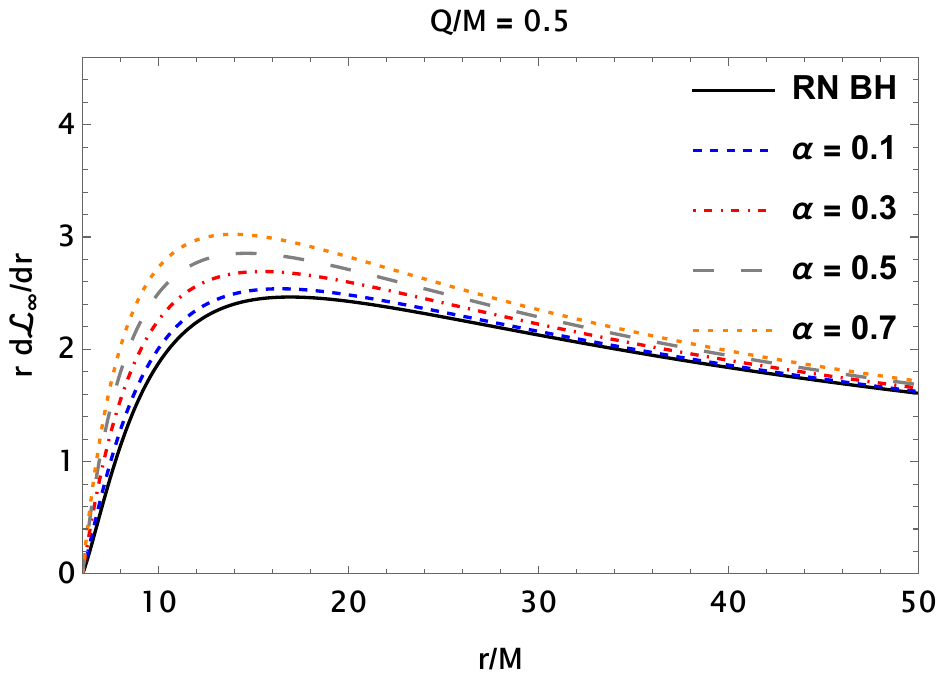}
    \caption{The plot shows the radial profile of the accretion disk's differential luminosity on the order of $10^{-2}$.}
    \label{fig:luminosity}
\end{figure}

\section{The accretion disk around a 4D charged Einstein-Gauss-Bonnet black hole and its radiative efficiency}\label{Sec:Model}

In this section, we consider the accretion disk around a charged black hole in 4D EGB gravity using the Novikov-Thorne model. According to this model, we assume that the accretion disk is optically thick and geometrically thin around the black hole. The point to note is that the accretion disk can be extended horizontally as its vertical size is considered thin enough. With this in view, the accretion disk can be characterized by two size parameters, such as the height $h$ and the radius $r$ of the disk. For these two parameters, $h/r \ll 1$ is then satisfied well as the disk's radius is much more larger in the horizontal direction. Hence, this property of the this accretion disk can cause the vertical entropy and the pressure gradients to be negligible for a gas and dust/particle in the disk. However, the accretion disk becomes hot enough and generates heat as a result of the dynamical friction, which can be released as the thermal radiation on the surface of the accretion disk starting from its inner edge located at the innermost stable circular orbit (ISCO) around the black hole. 

Keeping the above in mind we turn to the analysis of the accretions disc. To that end, we begin to examine the accretion disk's bolometric luminosity, which is defined by  \cite{Bokhari20,Rayimbaev-Shaymatov21a}
\begin{eqnarray}
\mathcal{L}_{bol}=\eta \dot{M}c^2\, ,
\end{eqnarray}
where $\dot{M}$ is the growth rate of matter that gets absorbed by the black hole and $\eta$ the energy efficiency of the accretion disk. From an astrophysical point of view, observing bolometric luminosity is still a complicated process due to black hole parameters and its geometry. Therefore, it is valuable to measure the bolometric luminosity using theoretical analysis and models since energy efficiency can be applied for explaining important aspects of the accretion process. It is envisaged that the remaining of mass-accreting matter is converted into the electromagnetic radiation that is emitted from the black hole. This is what we further define as the disk's energy efficiency extracted from black hole due to the infalling matter. This energy efficiency can be represented by the radiation rate of the photon energy emitted from the disk surface; see details in~\cite{Novikov1973,1974ApJ...191..499P}. We then proceed to determine the energy efficiency through the energy measured at the innermost stable circular orbits (ISCO), usually referred to as the emitted photons from the disk that travel out to infinity as radiation. This is defined by  
\begin{equation}
    \eta=1-\mathcal{E}_{ISCO}\, .
\end{equation}
From the above relation, we can determine the energy efficiency extracted from the accretion disc around the black hole. To do this, we need to determine the energy measured at the ISCO first.  
\begin{table}[htb]
\begin{center}
\caption{Table shows the ISCO radii of the neutral particle and the accretion disk's radiative energy efficiency for the various combinations of the GB coupling parameter $\alpha$ and the black hole charge $Q$.}\label{Tab1}
\resizebox{.48\textwidth}{!}
{
\begin{tabular}{l l|l l |l l| l l  }
 \hline \hline
 & &   $\alpha=0$  & &$\alpha=0.1$  & &$\alpha=0.3$    
 \\
\cline{3-5}\cline{6-8}
& $Q$    & $r_{ISCO}$ & $\eta\, [\%]$   & $r_{ISCO}$ & $\eta\, [\%]$ & $r_{ISCO}$ & $\eta\, [\%]$   \\
\hline
& 0.0   & 6.0 & 5.7191  & 5.93782  & 5.76376 & 5.80644  &5.85975  \\
& 0.2   & 5.94014 & 5.88378  & 5.87732  & 5.93054 & 5.74439  &6.03121 \\
& 0.4     & 5.75662 & 6.16915 & 5.69019  & 6.22239 & 5.5488 & 6.33783  \\
& 0.6       & 5.43249 & 6.62035 & 5.35783  &6.68781 & 5.19673 & 6.83654  \\
& 0.8       & 4.92279 & 7.35278 & 4.82951  &7.4542 & 4.6207 & 7.68743  \\
 \hline \hline
\end{tabular}
}
\end{center}
\end{table}

\begin{table}[htb]
\begin{center}
\caption{Table shows the ISCO radii of the charged particle and the accretion disk's radiative energy efficiency for the various combinations of the GB coupling parameter $\alpha$ and the black hole charge $Q$.}\label{Tab2}
\resizebox{.48\textwidth}{!}
{
\begin{tabular}{l l|l l |l l| l l  }
 \hline \hline
 & &   $\alpha=0$  & &$\alpha=0.1$  & &$\alpha=0.3$    
 \\
\cline{3-5}\cline{6-8}
& $Q$    & $r_{ISCO}$ & $\eta\, [\%]$   & $r_{ISCO}$ & $\eta\, [\%]$ & $r_{ISCO}$ & $\eta\, [\%]$   \\
\hline
%
%
& 0.2   & 5.93957 & 5.77232  & 5.87608  & 5.81878 & 5.74167  &5.91889 \\
& 0.4     & 5.75278 & 5.943 & 5.6849  & 5.99568 & 5.54023 & 6.11014  \\
& 0.6       & 5.41984 & 6.27224 & 5.34256  &6.33904 & 5.17532 & 6.48693  \\
& 0.8       & 4.89077 & 6.87159 & 4.79238  &6.9733 & 4.5867 & 7.21123  \\
 \hline \hline
\end{tabular}
}
\end{center}
\end{table}
%

Therefore, we now focus on the particle motion around BH in 4D EGB gravity. The effective potential for radial motion of a test particle with mass $m$ in the equatorial plane of a static black hole is written in general form \cite{Atamurotov21JCAP} 
\begin{eqnarray}
V^{\pm}_{\rm eff}(r)=q A_t \pm \sqrt{f(r)\left(1+\frac{{\cal L}^2}{r^2\sin^2\theta}\right)}\, ,
\end{eqnarray}
which defines the radial motion of the charged test particles around the black hole in 4D EGB gravity. It is clear that the effective potential has two parts, i.e., Column and gravitational interaction parts. Also, the effective potential consists of two different solutions and it keeps its symmetry under the transformation, such as $qQ \to -qQ$: $V^+_{\rm eff} \to -V^-_{\rm eff}$ and $V^-_{\rm eff} \to -V^+_{\rm eff}$. It is to be emphasized that we further focus on the positive one, $V^+_{\rm eff}$ that reflects physically meaningful timelike motion. Based on the effective potential, one can easily proceed to determine the ISCO parameters, such as $r_{ISCO}$, $\mathcal{E}_{ISCO}$ and $\mathcal{L}_{ISCO}$. To find these quantities, we impose the following conditions  
\begin{equation}\label{Eq:Veff_con}
    V_{\rm eff}(r)=0, \hspace{0.5cm} V'_{\rm eff}(r)=0 \hspace{0.5cm}\mbox{and} \hspace{0.5cm}V''_{\rm eff}(r)=0\, .
\end{equation}
All results associated with the ISCO radii for various combinations of the GB coupling parameter and black hole charge are tabulated in Table~\ref{Tab1} and \ref{Tab2}. As seen in Table~\ref{Tab1} and \ref{Tab2}, the ISCO radius decreases as $\alpha$ and $Q$ increases. Here, the GB coupling parameter can be interpreted as a gravitational repulsive charge, thus resulting in the gravity weakening in the background geometry \cite{Shaymatov20egb}. Therefore, it influences the radiative efficiency of the accretion disk around the black hole and it increases with increasing the GB coupling parameter. It is clear that the radiative efficiency is significantly enhanced due to the combined effects of the GB coupling $\alpha$ parameter and the black hole charge, compared to the the Schwarzschild and the RN black hole case in Einstein gravity \cite{Kurmanov_2022}. Interestingly, it can be observed that the radiative energy efficiency of the accretion disk is higher for neutral particles compared to charged particles. This is due to the Coulomb interaction, which allows charged particles to retain their energy in the accretion disk.       

\section{The accretion disk's radiative properties around a 4D charged Einstein-Gauss-Bonnet black hole}\label{section4}

Here, we investigate the accretion disk's radiative properties, such as the flux created in the  accretion disk around the charged black hole in 4D EGB gravity. It is worth noting that in most cases, accretion disks can be found around massive black holes. Any matter that enters the accretion disk can follow a trajectory in a spiral form within the disk, which is formed by gas and dust/particle orbiting around a black hole. Therefore, the gas and particle within the disk can rub and collide with each other as they move in a turbulent flow, causing frictional heating that is released as radiation energy. As a result of this process, the angular momentum of the gas and dust/particle decreases, causing them to drift inward towards the black hole. As they orbit closer to the black hole, their velocity increases, leading to an increase in frictional energy. Within this process, more energy is radiated away, resulting in the accretion disk becoming hot enough to emit X-rays in the close vicinity of the black hole. The main objective of this study is to investigate how the GB coupling parameter and the black hole charge can affect the accretion disk radiation. This may cause the gas and dust within the accretion disk to become highly ionized around the charged black hole, leading to the emission of exceptionally high-energy radiation such as X-rays that can be observed. Studying the accretion disk radiation around the charged black hole in 4D EGB gravity and analyzing the effects of its parameters on the disk can provide valuable insights into the properties of the disk and help us gain a deeper understanding of remarkable aspects of the black hole geometry. Following to~\cite{Novikov1973,Shakura:1972te,Thorne:1974ve}, we determine the flux of electromagnetic radiation via the following  equation
\begin{equation}\label{flux}
    \mathcal{F}(r)=-\dfrac{\dot{M_0}}{4 \pi \sqrt{g}}\dfrac{\Omega_{,r}}{(E-\Omega L)^2} \int _{r_{ISCO}}^r (E-\Omega L) L_{,r} d r\ , 
\end{equation}
where we denote $g$ as the determinant of the three-dimensional subspace with ($t, r , \phi$), i.e., $\sqrt{g}=\sqrt{-g_{tt}g_{rr}g_{\phi \phi}}$. From the flux equation given above, $\mathcal{F}(r)$ is sensitive to the disk mass accretion rate, $\dot{M_0}$ which remains unknown. For further analysis, we shall for simplicity choose $\dot{M_0}$. Here, the angular velocity (i.e., the Keplerian frequency) of the charged test particle is defined by~\cite{Shaymatov2022,Shaymatov23EPJP}
\begin{eqnarray}
   && \Omega^2=-\frac{g_{tt,r}}{g_{\phi\phi,r}}-\frac{2q^2}{m^2}\frac{g_{\phi\phi}A^2_{t,r}}{g_{\phi\phi,r}}+\frac{1}{g^2_{\phi\phi,r}} \Big[\frac{4q^4}{m^4}g^2_{\phi\phi}A^4_{t,r}+\nonumber \\
   && + \frac{4q^2}{m^2}A^2_{t,r}g_{\phi\phi,r}(g_{\phi\phi}g_{tt,r}-g_{tt}g_{\phi\phi,r})\Big]^{\frac{1}{2}}\, .
\end{eqnarray}
The angular velocity reduces to $\Omega^2_0=-g_{tt,r}/g_{\phi\phi,r}$ in the limit of $q\to 0$, referred to as the Keplerian frequency for a neutral particle. By imposing the conditions Eq.~(\ref{Eq:Veff_con}) for the energy and angular momentum, the flux of the electromagnetic radiation, $\mathcal{F}(r)$, can be determined explicitly. We now turn to analyze the flux numerically as deriving its analytical form is complicated. We show the radial dependence of the flux of electromagnetic radiation in Fig.~\ref{fig:flux}. The left and the right panels depict the impact of the GB coupling parameter and the black hole charge on the electromagnetic radiation flux, respectively. As can be seen from Fig.~\ref{fig:flux}, the shape of the electromagnetic radiation flux of the accretion disk is shifted upward towards larger values due to the impact of the GB coupling parameter $\alpha$, compared to the RN black hole case in Einstein gravity. The combined effects of the GB coupling parameter and the black hole charge can also enhance the flux significantly, as seen in the right panel of Fig.~\ref{fig:flux}. The electromagnetic radiation flux can reach its maximum value in the accretion disk as expected under the influences of $\alpha$ and $Q$. We are able to define the black body radiation flux as $\mathcal{F}(r)=\sigma T^4$, together with the Stefan-Boltzmann constant, $\sigma$. In Fig.~\ref{temperature1}, we show the radial profile of the accretion disk temperature as the disk energy. As seen in Fig.~Fig.~\ref{temperature1}, the maximum of the temperature profile corresponds to high temperature in the accretion disk around the black hole. The disk temperature increases as $\alpha$ increases, but it decreases accordingly at larger distances from the disk. To be more representative we show the density plot in Fig.~\ref{temperature2}, addressing the parameter dependence on the accretion disk temperature. In the top row of Fig.~\ref{temperature2}, the density plots depict how the disk temperature changes with the change in the GB coupling parameter and black hole charge. As observed from the density plot, the accretion disk temperature distribution starts to become hot enough from the inner edge of the disk and extends to its outer edge, as indicated by the red regions corresponding to the maximum temperature, as shown in the bottom panel of Fig.~\ref{temperature2}. 

Additionally, we now study the differential luminosity, which is one of key quantities and helps to reveals the accretion disk's nature~\cite{Novikov1973,Shakura:1972te,Thorne:1974ve}. The differential luminosity is written as follows:
\begin{equation}
    \dfrac{d \mathcal{L}_{\infty}}{d \ln{r}}=4 \pi r \sqrt{g} E \mathcal{F}(r)\, .
\end{equation}
It can be assumed that the radiation emission is regarded as the radiation of black body, thus allowing the spectral luminosity $\mathcal{L}_{\nu,\infty}$ to be defined by the radiation frequency variable $\nu$ at infinity~\cite{Boshkayev:2020kle,sym14091765,Alloqulov24CPC}
\begin{equation}\label{luminosity2}
    \nu \mathcal{L}_{\nu,\infty}=\dfrac{60}{\pi^3} \int_{r_{ISCO}}^{\infty} \dfrac{\sqrt{g} E}{M_T^2}\dfrac{(u^t y)^4}{\exp\Big[{\dfrac{u^t y}{(M_T^2 \mathcal{F})^{1/4}}}\Big]-1}\, ,
\end{equation}
where we have defined $y=h \nu /k T_{\star}$ with the characteristic temperature $T_{\star}$. Here, $k$ and $h$ respectively refer to the Planck and Boltzmann constants, while $M_T$ to the total mass. The characteristic temperature can be determined via the Stefan-Boltzmann law, and it is then given by 
\begin{equation}
  \sigma T_{\star}= \dfrac{\dot{M}_0}{4 \pi M_T^2}\, ,   
\end{equation}
with the Stefan-Boltzmann constant $\sigma$. We examine the differential luminosity and show its radial dependence in Fig.~\ref{fig:luminosity}. Similarly to what is observed in the electromagnetic flux, the differential luminosity can have a similar behavior as a consequence of the GB coupling parameter's impact. It increases in the accretion disk with the increasing of $\alpha$. Taking all into account, we can deduce that the accretion disk parameters are more sensitive to the coupling parameter $\alpha$ in EGB gravity case, compared to the RN black hole in Einstein's case.

\section{Conclusions }\label{Sec:Summary}

In this paper, we considered a charged black hole solution in 4D EGB gravity and investigated the efficiency of electric Penrose process and the accretion disk radiation properties around this black hole spacetime. We showed that the efficiency of the energy extraction from the black hole via the electric PP increases with the increasing black hole charge, but it is enhanced by the effect of positive values of the GB coupling parameter $\alpha$ for small values of black hole charge. However, the energy efficiency is strongly enhanced with increasing black hole charge that causes the Coulomb force interaction to increase. 

Accretion disks around black holes are considered the primary source of information regarding gravity and the surrounding geometry in the strong field regime. We also examined the radiative energy efficiency of the accretion disk. To be more quantitative, we estimated the radiative efficiency of the accretion disk using the ISCO energy of the charged particles for various possible cases and showed that the radiative efficiency increases as the ISCO radius becomes closer to the black hole as a consequence of an increase in $\alpha$ and $Q$; as seen in Table~\ref{Tab1} and \ref{Tab2}. Interestingly, it was found that the radiative efficiency is higher for neutral particles compared to charged particles. This occurs because the Coulomb interaction can allow charged particles to retain their energy in the accretion disk. 

Further, we studied the accretion disk's radiative properties and examined the effects of 4D EGB gravity to provide valuable insights into the properties of the disk. We found that the electromagnetic radiation flux of the accretion disk is significantly enhanced due to the impact of the positive GB coupling parameter $\alpha$, compared to the RN black hole case in Einstein gravity. It was also shown that the flux increase under the combined effects of the GB coupling parameter $\alpha$ and the black hole charge $Q$. Additionally, we examined the accretion disk temperature as a function of $\alpha$ and $Q$ and showed that it starts to increase and reach its maximum due to the increase in $\alpha$, but it decreases at larger distances from the disk. Based on the density plot, we also showed that the temperature distribution of the accretion disk starts to become hot in the accretion disk as a result of the combined effects of the coupling parameter and black hole charge. Finally, we studied the differential luminosity and demonstrated that it begins to increase with the rise in the value of the GB coupling parameter.

We examined the distinct future and effect of the GB coupling parameter and black hole charge on the efficiency of energy extraction and properties of the accretion disk. Based on the results, we found that the 4D EGB black hole is more efficient for the accretion disk as compared to the RN black hole in Einstein gravity case. These theoretical results may help assess the validity of 4D EGB black holes in explaining the distinct departures from Einstein gravity and astrophysical observations regarding the accretion disk radiations.

\section*{ACKNOWLEDGEMENT}

This work is supported by the National Natural Science Foundation of China under Grant No. 11675143 and the National Key Research and Development Program of China under Grant No. 2020YFC2201503. 

\bibliography{gravreferences,ref}

\begin{thebibliography}{79}%
\makeatletter
\providecommand \@ifxundefined [1]{%
 \@ifx{#1\undefined}
}%
\providecommand \@ifnum [1]{%
 \ifnum #1\expandafter \@firstoftwo
 \else \expandafter \@secondoftwo
 \fi
}%
\providecommand \@ifx [1]{%
 \ifx #1\expandafter \@firstoftwo
 \else \expandafter \@secondoftwo
 \fi
}%
\providecommand \natexlab [1]{#1}%
\providecommand \enquote  [1]{``#1''}%
\providecommand \bibnamefont  [1]{#1}%
\providecommand \bibfnamefont [1]{#1}%
\providecommand \citenamefont [1]{#1}%
\providecommand \href@noop [0]{\@secondoftwo}%
\providecommand \href [0]{\begingroup \@sanitize@url \@href}%
\providecommand \@href[1]{\@@startlink{#1}\@@href}%
\providecommand \@@href[1]{\endgroup#1\@@endlink}%
\providecommand \@sanitize@url [0]{\catcode `\\12\catcode `\$12\catcode
  `\&12\catcode `\#12\catcode `\^12\catcode `\_12\catcode `\%12\relax}%
\providecommand \@@startlink[1]{}%
\providecommand \@@endlink[0]{}%
\providecommand \url  [0]{\begingroup\@sanitize@url \@url }%
\providecommand \@url [1]{\endgroup\@href {#1}{\urlprefix }}%
\providecommand \urlprefix  [0]{URL }%
\providecommand \Eprint [0]{\href }%
\providecommand \doibase [0]{https://doi.org/}%
\providecommand \selectlanguage [0]{\@gobble}%
\providecommand \bibinfo  [0]{\@secondoftwo}%
\providecommand \bibfield  [0]{\@secondoftwo}%
\providecommand \translation [1]{[#1]}%
\providecommand \BibitemOpen [0]{}%
\providecommand \bibitemStop [0]{}%
\providecommand \bibitemNoStop [0]{.\EOS\space}%
\providecommand \EOS [0]{\spacefactor3000\relax}%
\providecommand \BibitemShut  [1]{\csname bibitem#1\endcsname}%
\let\auto@bib@innerbib\@empty
\bibitem [{\citenamefont {{Dadhich}}\ \emph {et~al.}(2012)\citenamefont
  {{Dadhich}}, \citenamefont {{Ghosh}},\ and\ \citenamefont
  {{Jhingan}}}]{Dadhich12c}%
  \BibitemOpen
  \bibfield  {author} {\bibinfo {author} {\bibfnamefont {N.}~\bibnamefont
  {{Dadhich}}}, \bibinfo {author} {\bibfnamefont {S.~G.}\ \bibnamefont
  {{Ghosh}}},\ and\ \bibinfo {author} {\bibfnamefont {S.}~\bibnamefont
  {{Jhingan}}},\ }\bibfield  {title} {\bibinfo {title} {{The Lovelock gravity
  in the critical spacetime dimension}},\ }\href
  {https://doi.org/10.1016/j.physletb.2012.03.084} {\bibfield  {journal}
  {\bibinfo  {journal} {Phys. Lett. B}\ }\textbf {\bibinfo {volume} {711}},\
  \bibinfo {pages} {196} (\bibinfo {year} {2012})},\ \Eprint
  {https://arxiv.org/abs/1202.4575} {arXiv:1202.4575 [gr-qc]} \BibitemShut
  {NoStop}%
\bibitem [{\citenamefont {{Lovelock}}(1971)}]{Lovelock1971}%
  \BibitemOpen
  \bibfield  {author} {\bibinfo {author} {\bibfnamefont {D.}~\bibnamefont
  {{Lovelock}}},\ }\bibfield  {title} {\bibinfo {title} {{The Einstein Tensor
  and Its Generalizations}},\ }\href {https://doi.org/10.1063/1.1665613}
  {\bibfield  {journal} {\bibinfo  {journal} {J. Math. Phys.}\ }\textbf
  {\bibinfo {volume} {12}},\ \bibinfo {pages} {498} (\bibinfo {year}
  {1971})}\BibitemShut {NoStop}%
\bibitem [{\citenamefont {{Glavan}}\ and\ \citenamefont
  {{Lin}}(2020)}]{Glavan20prl}%
  \BibitemOpen
  \bibfield  {author} {\bibinfo {author} {\bibfnamefont {D.}~\bibnamefont
  {{Glavan}}}\ and\ \bibinfo {author} {\bibfnamefont {C.}~\bibnamefont
  {{Lin}}},\ }\bibfield  {title} {\bibinfo {title} {{Einstein-Gauss-Bonnet
  Gravity in Four-Dimensional Spacetime}},\ }\href
  {https://doi.org/10.1103/PhysRevLett.124.081301} {\bibfield  {journal}
  {\bibinfo  {journal} {Phys. Rev. Lett.}\ }\textbf {\bibinfo {volume} {124}},\
  \bibinfo {eid} {081301} (\bibinfo {year} {2020})},\ \Eprint
  {https://arxiv.org/abs/1905.03601} {arXiv:1905.03601 [gr-qc]} \BibitemShut
  {NoStop}%
\bibitem [{\citenamefont {{G{\"u}rses}}\ \emph {et~al.}(2020)\citenamefont
  {{G{\"u}rses}}, \citenamefont {{{\c{S}}i{\textcommabelow s}man}},\ and\
  \citenamefont {{Tekin}}}]{Gurses20egb}%
  \BibitemOpen
  \bibfield  {author} {\bibinfo {author} {\bibfnamefont {M.}~\bibnamefont
  {{G{\"u}rses}}}, \bibinfo {author} {\bibfnamefont {T.~{\c{c}}.}\ \bibnamefont
  {{{\c{S}}i{\textcommabelow s}man}}},\ and\ \bibinfo {author} {\bibfnamefont
  {B.}~\bibnamefont {{Tekin}}},\ }\bibfield  {title} {\bibinfo {title} {{Is
  there a novel Einstein-Gauss-Bonnet theory in four dimensions?}},\ }\href
  {https://doi.org/10.1140/epjc/s10052-020-8200-7} {\bibfield  {journal}
  {\bibinfo  {journal} {Eur. Phys. J. C}\ }\textbf {\bibinfo {volume} {80}},\
  \bibinfo {eid} {647} (\bibinfo {year} {2020})},\ \Eprint
  {https://arxiv.org/abs/2004.03390} {arXiv:2004.03390 [gr-qc]} \BibitemShut
  {NoStop}%
\bibitem [{\citenamefont {{Mahapatra}}(2020)}]{Mahapatra20egb}%
  \BibitemOpen
  \bibfield  {author} {\bibinfo {author} {\bibfnamefont {S.}~\bibnamefont
  {{Mahapatra}}},\ }\bibfield  {title} {\bibinfo {title} {{A note on the total
  action of 4D Gauss-Bonnet theory}},\ }\href
  {https://doi.org/10.1140/epjc/s10052-020-08568-6} {\bibfield  {journal}
  {\bibinfo  {journal} {Eur. Phys. J. C}\ }\textbf {\bibinfo {volume} {80}},\
  \bibinfo {eid} {992} (\bibinfo {year} {2020})},\ \Eprint
  {https://arxiv.org/abs/2004.09214} {arXiv:2004.09214 [gr-qc]} \BibitemShut
  {NoStop}%
\bibitem [{\citenamefont {{Hennigar}}\ \emph
  {et~al.}(2020{\natexlab{a}})\citenamefont {{Hennigar}}, \citenamefont
  {{Kubiz{\v{n}}{\'a}k}}, \citenamefont {{Mann}},\ and\ \citenamefont
  {{Pollack}}}]{Hennigar20egb}%
  \BibitemOpen
  \bibfield  {author} {\bibinfo {author} {\bibfnamefont {R.~A.}\ \bibnamefont
  {{Hennigar}}}, \bibinfo {author} {\bibfnamefont {D.}~\bibnamefont
  {{Kubiz{\v{n}}{\'a}k}}}, \bibinfo {author} {\bibfnamefont {R.~B.}\
  \bibnamefont {{Mann}}},\ and\ \bibinfo {author} {\bibfnamefont
  {C.}~\bibnamefont {{Pollack}}},\ }\bibfield  {title} {\bibinfo {title} {{On
  taking the D {\textrightarrow} 4 limit of Gauss-Bonnet gravity: theory and
  solutions}},\ }\href {https://doi.org/10.1007/JHEP07(2020)027} {\bibfield
  {journal} {\bibinfo  {journal} {JHEP}\ }\textbf {\bibinfo {volume}
  {2020}}\bibfield  {number} {\bibinfo  {number} { (7)},\ \bibinfo {eid}
  {27}},\ }\Eprint {https://arxiv.org/abs/2004.09472} {arXiv:2004.09472
  [gr-qc]} \BibitemShut {NoStop}%
\bibitem [{\citenamefont {{Liu}}\ \emph {et~al.}(2021)\citenamefont {{Liu}},
  \citenamefont {{Zhu}},\ and\ \citenamefont {{Wu}}}]{Liu20egb}%
  \BibitemOpen
  \bibfield  {author} {\bibinfo {author} {\bibfnamefont {C.}~\bibnamefont
  {{Liu}}}, \bibinfo {author} {\bibfnamefont {T.}~\bibnamefont {{Zhu}}},\ and\
  \bibinfo {author} {\bibfnamefont {Q.}~\bibnamefont {{Wu}}},\ }\bibfield
  {title} {\bibinfo {title} {{Thin accretion disk around a four-dimensional
  Einstein-Gauss-Bonnet black hole}},\ }\href
  {https://doi.org/10.1088/1674-1137/abc16c} {\bibfield  {journal} {\bibinfo
  {journal} {Chin. Phys. C}\ }\textbf {\bibinfo {volume} {45}},\ \bibinfo {eid}
  {015105} (\bibinfo {year} {2021})},\ \Eprint
  {https://arxiv.org/abs/2004.01662} {arXiv:2004.01662 [gr-qc]} \BibitemShut
  {NoStop}%
\bibitem [{\citenamefont {{Guo}}\ and\ \citenamefont {{Li}}(2020)}]{Guo20egb}%
  \BibitemOpen
  \bibfield  {author} {\bibinfo {author} {\bibfnamefont {M.}~\bibnamefont
  {{Guo}}}\ and\ \bibinfo {author} {\bibfnamefont {P.-C.}\ \bibnamefont
  {{Li}}},\ }\bibfield  {title} {\bibinfo {title} {{Innermost stable circular
  orbit and shadow of the 4D Einstein-Gauss-Bonnet black hole}},\ }\href
  {https://doi.org/10.1140/epjc/s10052-020-8164-7} {\bibfield  {journal}
  {\bibinfo  {journal} {Eur. Phys. J. C}\ }\textbf {\bibinfo {volume} {80}},\
  \bibinfo {eid} {588} (\bibinfo {year} {2020})},\ \Eprint
  {https://arxiv.org/abs/2003.02523} {arXiv:2003.02523 [gr-qc]} \BibitemShut
  {NoStop}%
\bibitem [{\citenamefont {{Wei}}\ and\ \citenamefont {{Liu}}(2020)}]{Wei20egb}%
  \BibitemOpen
  \bibfield  {author} {\bibinfo {author} {\bibfnamefont {S.-W.}\ \bibnamefont
  {{Wei}}}\ and\ \bibinfo {author} {\bibfnamefont {Y.-X.}\ \bibnamefont
  {{Liu}}},\ }\bibfield  {title} {\bibinfo {title} {{Testing the nature of
  Gauss-Bonnet gravity by four-dimensional rotating black hole shadow}},\
  }\href@noop {} {\bibfield  {journal} {\bibinfo  {journal} {arXiv e-prints}\ }
  (\bibinfo {year} {2020})},\ \Eprint {https://arxiv.org/abs/2003.07769}
  {arXiv:2003.07769 [gr-qc]} \BibitemShut {NoStop}%
\bibitem [{\citenamefont {{Kumar}}\ and\ \citenamefont
  {{Ghosh}}(2020)}]{Kumar20egb}%
  \BibitemOpen
  \bibfield  {author} {\bibinfo {author} {\bibfnamefont {R.}~\bibnamefont
  {{Kumar}}}\ and\ \bibinfo {author} {\bibfnamefont {S.~G.}\ \bibnamefont
  {{Ghosh}}},\ }\bibfield  {title} {\bibinfo {title} {{Rotating black holes in
  4D Einstein-Gauss-Bonnet gravity and its shadow}},\ }\href
  {https://doi.org/10.1088/1475-7516/2020/07/053} {\bibfield  {journal}
  {\bibinfo  {journal} {JCAP}\ }\textbf {\bibinfo {volume} {2020}}\bibfield
  {number} {\bibinfo  {number} { (7)},\ \bibinfo {eid} {053}},\ }\Eprint
  {https://arxiv.org/abs/2003.08927} {arXiv:2003.08927 [gr-qc]} \BibitemShut
  {NoStop}%
\bibitem [{\citenamefont {{Konoplya}}\ and\ \citenamefont
  {{Zinhailo}}(2020)}]{Konoplya20egb}%
  \BibitemOpen
  \bibfield  {author} {\bibinfo {author} {\bibfnamefont {R.~A.}\ \bibnamefont
  {{Konoplya}}}\ and\ \bibinfo {author} {\bibfnamefont {A.~F.}\ \bibnamefont
  {{Zinhailo}}},\ }\bibfield  {title} {\bibinfo {title} {{Quasinormal modes,
  stability and shadows of a black hole in the 4D Einstein-Gauss-Bonnet
  gravity}},\ }\href {https://doi.org/10.1140/epjc/s10052-020-08639-8}
  {\bibfield  {journal} {\bibinfo  {journal} {Eur. Phys. J. C}\ }\textbf
  {\bibinfo {volume} {80}},\ \bibinfo {eid} {1049} (\bibinfo {year} {2020})},\
  \Eprint {https://arxiv.org/abs/2003.01188} {arXiv:2003.01188 [gr-qc]}
  \BibitemShut {NoStop}%
\bibitem [{\citenamefont {{Churilova}}(2020)}]{Churilova20egb}%
  \BibitemOpen
  \bibfield  {author} {\bibinfo {author} {\bibfnamefont {M.~S.}\ \bibnamefont
  {{Churilova}}},\ }\bibfield  {title} {\bibinfo {title} {{Quasinormal modes of
  the Dirac field in the novel 4D Einstein-Gauss-Bonnet gravity}},\ }\href@noop
  {} {\bibfield  {journal} {\bibinfo  {journal} {arXiv e-prints}\ } (\bibinfo
  {year} {2020})},\ \Eprint {https://arxiv.org/abs/2004.00513}
  {arXiv:2004.00513 [gr-qc]} \BibitemShut {NoStop}%
\bibitem [{\citenamefont {{Malafarina}}\ \emph {et~al.}(2020)\citenamefont
  {{Malafarina}}, \citenamefont {{Toshmatov}},\ and\ \citenamefont
  {{Dadhich}}}]{Malafarina20egb}%
  \BibitemOpen
  \bibfield  {author} {\bibinfo {author} {\bibfnamefont {D.}~\bibnamefont
  {{Malafarina}}}, \bibinfo {author} {\bibfnamefont {B.}~\bibnamefont
  {{Toshmatov}}},\ and\ \bibinfo {author} {\bibfnamefont {N.}~\bibnamefont
  {{Dadhich}}},\ }\bibfield  {title} {\bibinfo {title} {{Dust collapse in 4D
  Einstein-Gauss-Bonnet gravity}},\ }\href
  {https://doi.org/10.1016/j.dark.2020.100598} {\bibfield  {journal} {\bibinfo
  {journal} {Phys. Dark Universe}\ }\textbf {\bibinfo {volume} {30}},\ \bibinfo
  {eid} {100598} (\bibinfo {year} {2020})},\ \Eprint
  {https://arxiv.org/abs/2004.07089} {arXiv:2004.07089 [gr-qc]} \BibitemShut
  {NoStop}%
\bibitem [{\citenamefont {{Arag{\'o}n}}\ \emph {et~al.}(2020)\citenamefont
  {{Arag{\'o}n}}, \citenamefont {{B{\'e}car}}, \citenamefont {{Gonz{\'a}lez}},\
  and\ \citenamefont {{V{\'a}squez}}}]{Aragon20egb}%
  \BibitemOpen
  \bibfield  {author} {\bibinfo {author} {\bibfnamefont {A.}~\bibnamefont
  {{Arag{\'o}n}}}, \bibinfo {author} {\bibfnamefont {R.}~\bibnamefont
  {{B{\'e}car}}}, \bibinfo {author} {\bibfnamefont {P.~A.}\ \bibnamefont
  {{Gonz{\'a}lez}}},\ and\ \bibinfo {author} {\bibfnamefont {Y.}~\bibnamefont
  {{V{\'a}squez}}},\ }\bibfield  {title} {\bibinfo {title} {{Perturbative and
  nonperturbative quasinormal modes of 4D Einstein-Gauss-Bonnet black holes}},\
  }\href {https://doi.org/10.1140/epjc/s10052-020-8298-7} {\bibfield  {journal}
  {\bibinfo  {journal} {Eur. Phys. J. C}\ }\textbf {\bibinfo {volume} {80}},\
  \bibinfo {eid} {773} (\bibinfo {year} {2020})},\ \Eprint
  {https://arxiv.org/abs/2004.05632} {arXiv:2004.05632 [gr-qc]} \BibitemShut
  {NoStop}%
\bibitem [{\citenamefont {{Mansoori}}(2020)}]{Mansoori20egb}%
  \BibitemOpen
  \bibfield  {author} {\bibinfo {author} {\bibfnamefont {S.~A.~H.}\
  \bibnamefont {{Mansoori}}},\ }\bibfield  {title} {\bibinfo {title}
  {{Thermodynamic geometry of the novel 4-D Gauss Bonnet AdS Black Hole}},\
  }\href@noop {} {\bibfield  {journal} {\bibinfo  {journal} {arXiv e-prints}\ }
  (\bibinfo {year} {2020})},\ \Eprint {https://arxiv.org/abs/2003.13382}
  {arXiv:2003.13382 [gr-qc]} \BibitemShut {NoStop}%
\bibitem [{\citenamefont {{Abdujabbarov}}\ \emph {et~al.}(2020)\citenamefont
  {{Abdujabbarov}}, \citenamefont {{Rayimbaev}}, \citenamefont {{Turimov}},\
  and\ \citenamefont {{Atamurotov}}}]{Rayimbaev2020egb}%
  \BibitemOpen
  \bibfield  {author} {\bibinfo {author} {\bibfnamefont {A.}~\bibnamefont
  {{Abdujabbarov}}}, \bibinfo {author} {\bibfnamefont {J.}~\bibnamefont
  {{Rayimbaev}}}, \bibinfo {author} {\bibfnamefont {B.}~\bibnamefont
  {{Turimov}}},\ and\ \bibinfo {author} {\bibfnamefont {F.}~\bibnamefont
  {{Atamurotov}}},\ }\bibfield  {title} {\bibinfo {title} {{Dynamics of
  magnetized particles around 4-D Einstein Gauss-Bonnet black hole}},\ }\href
  {https://doi.org/10.1016/j.dark.2020.100715} {\bibfield  {journal} {\bibinfo
  {journal} {Phys. Dark Universe}\ }\textbf {\bibinfo {volume} {30}},\ \bibinfo
  {eid} {100715} (\bibinfo {year} {2020})}\BibitemShut {NoStop}%
\bibitem [{\citenamefont {{Ge}}\ and\ \citenamefont {{Sin}}(2020)}]{Ge20egb}%
  \BibitemOpen
  \bibfield  {author} {\bibinfo {author} {\bibfnamefont {X.-H.}\ \bibnamefont
  {{Ge}}}\ and\ \bibinfo {author} {\bibfnamefont {S.-J.}\ \bibnamefont
  {{Sin}}},\ }\bibfield  {title} {\bibinfo {title} {{Causality of black holes
  in 4-dimensional Einstein-Gauss-Bonnet-Maxwell theory}},\ }\href
  {https://doi.org/10.1140/epjc/s10052-020-8288-9} {\bibfield  {journal}
  {\bibinfo  {journal} {Eur. Phys. J. C}\ }\textbf {\bibinfo {volume} {80}},\
  \bibinfo {eid} {695} (\bibinfo {year} {2020})},\ \Eprint
  {https://arxiv.org/abs/2004.12191} {arXiv:2004.12191 [hep-th]} \BibitemShut
  {NoStop}%
\bibitem [{\citenamefont {{Chakraborty}}\ and\ \citenamefont
  {{Dadhich}}(2020)}]{Chakraborty20egb}%
  \BibitemOpen
  \bibfield  {author} {\bibinfo {author} {\bibfnamefont {S.}~\bibnamefont
  {{Chakraborty}}}\ and\ \bibinfo {author} {\bibfnamefont {N.}~\bibnamefont
  {{Dadhich}}},\ }\bibfield  {title} {\bibinfo {title} {{Limits on stellar
  structures in Lovelock theories of gravity}},\ }\href
  {https://doi.org/10.1016/j.dark.2020.100658} {\bibfield  {journal} {\bibinfo
  {journal} {Phys. Dark Universe}\ }\textbf {\bibinfo {volume} {30}},\ \bibinfo
  {eid} {100658} (\bibinfo {year} {2020})},\ \Eprint
  {https://arxiv.org/abs/2005.07504} {arXiv:2005.07504 [gr-qc]} \BibitemShut
  {NoStop}%
\bibitem [{\citenamefont {{Odintsov}}\ and\ \citenamefont
  {{Oikonomou}}(2020)}]{Odintsov20plb}%
  \BibitemOpen
  \bibfield  {author} {\bibinfo {author} {\bibfnamefont {S.~D.}\ \bibnamefont
  {{Odintsov}}}\ and\ \bibinfo {author} {\bibfnamefont {V.~K.}\ \bibnamefont
  {{Oikonomou}}},\ }\bibfield  {title} {\bibinfo {title} {{Swampland
  implications of GW170817-compatible Einstein-Gauss-Bonnet gravity}},\ }\href
  {https://doi.org/10.1016/j.physletb.2020.135437} {\bibfield  {journal}
  {\bibinfo  {journal} {Phys. Lett. B}\ }\textbf {\bibinfo {volume} {805}},\
  \bibinfo {eid} {135437} (\bibinfo {year} {2020})},\ \Eprint
  {https://arxiv.org/abs/2004.00479} {arXiv:2004.00479 [gr-qc]} \BibitemShut
  {NoStop}%
\bibitem [{\citenamefont {{Lin}}\ \emph {et~al.}(2020)\citenamefont {{Lin}},
  \citenamefont {{Yang}}, \citenamefont {{Wei}}, \citenamefont {{Wang}},\ and\
  \citenamefont {{Liu}}}]{Lin20egb}%
  \BibitemOpen
  \bibfield  {author} {\bibinfo {author} {\bibfnamefont {Z.-C.}\ \bibnamefont
  {{Lin}}}, \bibinfo {author} {\bibfnamefont {K.}~\bibnamefont {{Yang}}},
  \bibinfo {author} {\bibfnamefont {S.-W.}\ \bibnamefont {{Wei}}}, \bibinfo
  {author} {\bibfnamefont {Y.-Q.}\ \bibnamefont {{Wang}}},\ and\ \bibinfo
  {author} {\bibfnamefont {Y.-X.}\ \bibnamefont {{Liu}}},\ }\bibfield  {title}
  {\bibinfo {title} {{Equivalence of solutions between the four-dimensional
  novel and regularized EGB theories in a cylindrically symmetric spacetime}},\
  }\href {https://doi.org/10.1140/epjc/s10052-020-08612-5} {\bibfield
  {journal} {\bibinfo  {journal} {Eur. Phys. J. C}\ }\textbf {\bibinfo {volume}
  {80}},\ \bibinfo {eid} {1033} (\bibinfo {year} {2020})},\ \Eprint
  {https://arxiv.org/abs/2006.07913} {arXiv:2006.07913 [gr-qc]} \BibitemShut
  {NoStop}%
\bibitem [{\citenamefont {{Aoki}}\ \emph {et~al.}(2020)\citenamefont {{Aoki}},
  \citenamefont {{Gorji}},\ and\ \citenamefont {{Mukohyama}}}]{Aoki20egb}%
  \BibitemOpen
  \bibfield  {author} {\bibinfo {author} {\bibfnamefont {K.}~\bibnamefont
  {{Aoki}}}, \bibinfo {author} {\bibfnamefont {M.~A.}\ \bibnamefont
  {{Gorji}}},\ and\ \bibinfo {author} {\bibfnamefont {S.}~\bibnamefont
  {{Mukohyama}}},\ }\bibfield  {title} {\bibinfo {title} {{A consistent theory
  of D {\textrightarrow} 4 Einstein-Gauss-Bonnet gravity}},\ }\href
  {https://doi.org/10.1016/j.physletb.2020.135843} {\bibfield  {journal}
  {\bibinfo  {journal} {Phys. Lett. B}\ }\textbf {\bibinfo {volume} {810}},\
  \bibinfo {eid} {135843} (\bibinfo {year} {2020})},\ \Eprint
  {https://arxiv.org/abs/2005.03859} {arXiv:2005.03859 [gr-qc]} \BibitemShut
  {NoStop}%
\bibitem [{\citenamefont {{Shaymatov}}\ \emph {et~al.}(2020)\citenamefont
  {{Shaymatov}}, \citenamefont {{Vrba}}, \citenamefont {{Malafarina}},
  \citenamefont {{Ahmedov}},\ and\ \citenamefont
  {{Stuchl{\'\i}k}}}]{Shaymatov20egb}%
  \BibitemOpen
  \bibfield  {author} {\bibinfo {author} {\bibfnamefont {S.}~\bibnamefont
  {{Shaymatov}}}, \bibinfo {author} {\bibfnamefont {J.}~\bibnamefont {{Vrba}}},
  \bibinfo {author} {\bibfnamefont {D.}~\bibnamefont {{Malafarina}}}, \bibinfo
  {author} {\bibfnamefont {B.}~\bibnamefont {{Ahmedov}}},\ and\ \bibinfo
  {author} {\bibfnamefont {Z.}~\bibnamefont {{Stuchl{\'\i}k}}},\ }\bibfield
  {title} {\bibinfo {title} {{Charged particle and epicyclic motions around 4 D
  Einstein-Gauss-Bonnet black hole immersed in an external magnetic field}},\
  }\href {https://doi.org/10.1016/j.dark.2020.100648} {\bibfield  {journal}
  {\bibinfo  {journal} {Phys. Dark Universe}\ }\textbf {\bibinfo {volume}
  {30}},\ \bibinfo {eid} {100648} (\bibinfo {year} {2020})},\ \Eprint
  {https://arxiv.org/abs/2005.12410} {arXiv:2005.12410 [gr-qc]} \BibitemShut
  {NoStop}%
\bibitem [{\citenamefont {{Islam}}\ \emph {et~al.}(2020)\citenamefont
  {{Islam}}, \citenamefont {{Kumar}},\ and\ \citenamefont
  {{Ghosh}}}]{Islam20egb}%
  \BibitemOpen
  \bibfield  {author} {\bibinfo {author} {\bibfnamefont {S.~U.}\ \bibnamefont
  {{Islam}}}, \bibinfo {author} {\bibfnamefont {R.}~\bibnamefont {{Kumar}}},\
  and\ \bibinfo {author} {\bibfnamefont {S.~G.}\ \bibnamefont {{Ghosh}}},\
  }\bibfield  {title} {\bibinfo {title} {{Gravitational lensing by black holes
  in the 4D Einstein-Gauss-Bonnet gravity}},\ }\href
  {https://doi.org/10.1088/1475-7516/2020/09/030} {\bibfield  {journal}
  {\bibinfo  {journal} {JCAP}\ }\textbf {\bibinfo {volume} {2020}}\bibfield
  {number} {\bibinfo  {number} { (9)},\ \bibinfo {eid} {030}},\ }\Eprint
  {https://arxiv.org/abs/2004.01038} {arXiv:2004.01038 [gr-qc]} \BibitemShut
  {NoStop}%
\bibitem [{\citenamefont {{Singh}}\ and\ \citenamefont
  {{Siwach}}(2020)}]{Singh20-egb}%
  \BibitemOpen
  \bibfield  {author} {\bibinfo {author} {\bibfnamefont {D.~V.}\ \bibnamefont
  {{Singh}}}\ and\ \bibinfo {author} {\bibfnamefont {S.}~\bibnamefont
  {{Siwach}}},\ }\bibfield  {title} {\bibinfo {title} {{Thermodynamics and P-v
  criticality of Bardeen-AdS black hole in 4D Einstein-Gauss-Bonnet gravity}},\
  }\href {https://doi.org/10.1016/j.physletb.2020.135658} {\bibfield  {journal}
  {\bibinfo  {journal} {Phys. Lett. B}\ }\textbf {\bibinfo {volume} {808}},\
  \bibinfo {eid} {135658} (\bibinfo {year} {2020})}\BibitemShut {NoStop}%
\bibitem [{\citenamefont {{Zhang}}\ \emph
  {et~al.}(2020{\natexlab{a}})\citenamefont {{Zhang}}, \citenamefont {{Wei}},\
  and\ \citenamefont {{Liu}}}]{Zhang20egb}%
  \BibitemOpen
  \bibfield  {author} {\bibinfo {author} {\bibfnamefont {Y.-P.}\ \bibnamefont
  {{Zhang}}}, \bibinfo {author} {\bibfnamefont {S.-W.}\ \bibnamefont {{Wei}}},\
  and\ \bibinfo {author} {\bibfnamefont {Y.-X.}\ \bibnamefont {{Liu}}},\
  }\bibfield  {title} {\bibinfo {title} {{Spinning Test Particle in
  Four-Dimensional Einstein-Gauss-Bonnet Black Holes}},\ }\href
  {https://doi.org/10.3390/universe6080103} {\bibfield  {journal} {\bibinfo
  {journal} {Universe}\ }\textbf {\bibinfo {volume} {6}},\ \bibinfo {pages}
  {103} (\bibinfo {year} {2020}{\natexlab{a}})},\ \Eprint
  {https://arxiv.org/abs/2003.10960} {arXiv:2003.10960 [gr-qc]} \BibitemShut
  {NoStop}%
\bibitem [{\citenamefont {{Zhang}}\ \emph
  {et~al.}(2020{\natexlab{b}})\citenamefont {{Zhang}}, \citenamefont {{Zhang}},
  \citenamefont {{Li}},\ and\ \citenamefont {{Guo}}}]{Zhang20aegb}%
  \BibitemOpen
  \bibfield  {author} {\bibinfo {author} {\bibfnamefont {C.-Y.}\ \bibnamefont
  {{Zhang}}}, \bibinfo {author} {\bibfnamefont {S.-J.}\ \bibnamefont
  {{Zhang}}}, \bibinfo {author} {\bibfnamefont {P.-C.}\ \bibnamefont {{Li}}},\
  and\ \bibinfo {author} {\bibfnamefont {M.}~\bibnamefont {{Guo}}},\ }\bibfield
   {title} {\bibinfo {title} {{Superradiance and stability of the novel 4D
  charged Einstein-Gauss-Bonnet black hole}},\ }\href@noop {} {\bibfield
  {journal} {\bibinfo  {journal} {arXiv e-prints}\ ,\ \bibinfo {eid}
  {arXiv:2004.03141}} (\bibinfo {year} {2020}{\natexlab{b}})},\ \Eprint
  {https://arxiv.org/abs/2004.03141} {arXiv:2004.03141 [gr-qc]} \BibitemShut
  {NoStop}%
\bibitem [{\citenamefont {{Atamurotov}}\ \emph {et~al.}(2021)\citenamefont
  {{Atamurotov}}, \citenamefont {{Shaymatov}}, \citenamefont {{Sheoran}},\ and\
  \citenamefont {{Siwach}}}]{Atamurotov21JCAP}%
  \BibitemOpen
  \bibfield  {author} {\bibinfo {author} {\bibfnamefont {F.}~\bibnamefont
  {{Atamurotov}}}, \bibinfo {author} {\bibfnamefont {S.}~\bibnamefont
  {{Shaymatov}}}, \bibinfo {author} {\bibfnamefont {P.}~\bibnamefont
  {{Sheoran}}},\ and\ \bibinfo {author} {\bibfnamefont {S.}~\bibnamefont
  {{Siwach}}},\ }\bibfield  {title} {\bibinfo {title} {{Charged black hole in
  4D Einstein-Gauss-Bonnet gravity: particle motion, plasma effect on weak
  gravitational lensing and centre-of-mass energy}},\ }\href
  {https://doi.org/10.1088/1475-7516/2021/08/045} {\bibfield  {journal}
  {\bibinfo  {journal} {JCAP}\ }\textbf {\bibinfo {volume} {2021}}\bibfield
  {number} {\bibinfo  {number} { (8)},\ \bibinfo {eid} {045}},\ }\Eprint
  {https://arxiv.org/abs/2105.02214} {arXiv:2105.02214 [gr-qc]} \BibitemShut
  {NoStop}%
\bibitem [{\citenamefont {{Narzilloev}}\ \emph {et~al.}(2021)\citenamefont
  {{Narzilloev}}, \citenamefont {{Shaymatov}}, \citenamefont {{Hussain}},
  \citenamefont {{Abdujabbarov}}, \citenamefont {{Ahmedov}},\ and\
  \citenamefont {{Bambi}}}]{Narzilloev21BTZ}%
  \BibitemOpen
  \bibfield  {author} {\bibinfo {author} {\bibfnamefont {B.}~\bibnamefont
  {{Narzilloev}}}, \bibinfo {author} {\bibfnamefont {S.}~\bibnamefont
  {{Shaymatov}}}, \bibinfo {author} {\bibfnamefont {I.}~\bibnamefont
  {{Hussain}}}, \bibinfo {author} {\bibfnamefont {A.}~\bibnamefont
  {{Abdujabbarov}}}, \bibinfo {author} {\bibfnamefont {B.}~\bibnamefont
  {{Ahmedov}}},\ and\ \bibinfo {author} {\bibfnamefont {C.}~\bibnamefont
  {{Bambi}}},\ }\bibfield  {title} {\bibinfo {title} {{Motion of particles and
  gravitational lensing around the (2+1)-dimensional BTZ black hole in
  Gauss-Bonnet gravity}},\ }\href
  {https://doi.org/10.1140/epjc/s10052-021-09617-4} {\bibfield  {journal}
  {\bibinfo  {journal} {Eur. Phys. J. C}\ }\textbf {\bibinfo {volume} {81}},\
  \bibinfo {eid} {849} (\bibinfo {year} {2021})},\ \Eprint
  {https://arxiv.org/abs/2109.02816} {arXiv:2109.02816 [gr-qc]} \BibitemShut
  {NoStop}%
\bibitem [{\citenamefont {{Mishra}}(2020)}]{Mishra20egb}%
  \BibitemOpen
  \bibfield  {author} {\bibinfo {author} {\bibfnamefont {A.~K.}\ \bibnamefont
  {{Mishra}}},\ }\bibfield  {title} {\bibinfo {title} {{Quasinormal modes and
  strong cosmic censorship in the regularised 4D Einstein-Gauss-Bonnet
  gravity}},\ }\href {https://doi.org/10.1007/s10714-020-02763-2} {\bibfield
  {journal} {\bibinfo  {journal} {Gen. Relativ. Gravit.}\ }\textbf {\bibinfo
  {volume} {52}},\ \bibinfo {eid} {106} (\bibinfo {year} {2020})},\ \Eprint
  {https://arxiv.org/abs/2004.01243} {arXiv:2004.01243 [gr-qc]} \BibitemShut
  {NoStop}%
\bibitem [{\citenamefont {{Yang}}\ \emph {et~al.}(2020)\citenamefont {{Yang}},
  \citenamefont {{Wan}}, \citenamefont {{Chen}}, \citenamefont {{Yang}},\ and\
  \citenamefont {{Wang}}}]{Yang20b}%
  \BibitemOpen
  \bibfield  {author} {\bibinfo {author} {\bibfnamefont {S.-J.}\ \bibnamefont
  {{Yang}}}, \bibinfo {author} {\bibfnamefont {J.-J.}\ \bibnamefont {{Wan}}},
  \bibinfo {author} {\bibfnamefont {J.}~\bibnamefont {{Chen}}}, \bibinfo
  {author} {\bibfnamefont {J.}~\bibnamefont {{Yang}}},\ and\ \bibinfo {author}
  {\bibfnamefont {Y.-Q.}\ \bibnamefont {{Wang}}},\ }\bibfield  {title}
  {\bibinfo {title} {{Weak cosmic censorship conjecture for the novel 4D
  charged Einstein-Gauss-Bonnet black hole with test scalar field and
  particle}},\ }\href {https://doi.org/10.1140/epjc/s10052-020-08511-9}
  {\bibfield  {journal} {\bibinfo  {journal} {Eur. Phys. J. C}\ }\textbf
  {\bibinfo {volume} {80}},\ \bibinfo {eid} {937} (\bibinfo {year} {2020})},\
  \Eprint {https://arxiv.org/abs/2004.07934} {arXiv:2004.07934 [gr-qc]}
  \BibitemShut {NoStop}%
\bibitem [{\citenamefont {{Ahmed}}\ \emph {et~al.}(2022)\citenamefont
  {{Ahmed}}, \citenamefont {{Shaymatov}},\ and\ \citenamefont
  {{Ahmedov}}}]{Ayyesha22PDU}%
  \BibitemOpen
  \bibfield  {author} {\bibinfo {author} {\bibfnamefont {A.~K.}\ \bibnamefont
  {{Ahmed}}}, \bibinfo {author} {\bibfnamefont {S.}~\bibnamefont
  {{Shaymatov}}},\ and\ \bibinfo {author} {\bibfnamefont {B.}~\bibnamefont
  {{Ahmedov}}},\ }\bibfield  {title} {\bibinfo {title} {{Weak cosmic censorship
  conjecture for the (2+1)-dimensional charged BTZ black hole in the
  Einstein-Gauss-Bonnet Gravity}},\ }\href
  {https://doi.org/10.1016/j.dark.2022.101082} {\bibfield  {journal} {\bibinfo
  {journal} {Phys. Dark Universe}\ }\textbf {\bibinfo {volume} {37}},\ \bibinfo
  {eid} {101082} (\bibinfo {year} {2022})},\ \Eprint
  {https://arxiv.org/abs/2207.01694} {arXiv:2207.01694 [gr-qc]} \BibitemShut
  {NoStop}%
\bibitem [{\citenamefont {{Donmez}}(2021)}]{Donmez2021egb}%
  \BibitemOpen
  \bibfield  {author} {\bibinfo {author} {\bibfnamefont {O.}~\bibnamefont
  {{Donmez}}},\ }\bibfield  {title} {\bibinfo {title} {{Bondi-Hoyle accretion
  around the non-rotating black hole in 4D Einstein-Gauss-Bonnet gravity}},\
  }\href {https://doi.org/10.1140/epjc/s10052-021-08923-1} {\bibfield
  {journal} {\bibinfo  {journal} {Eur. Phys. J. C}\ }\textbf {\bibinfo {volume}
  {81}},\ \bibinfo {eid} {113} (\bibinfo {year} {2021})},\ \Eprint
  {https://arxiv.org/abs/2011.04399} {arXiv:2011.04399 [gr-qc]} \BibitemShut
  {NoStop}%
\bibitem [{\citenamefont {{Fernandes}}(2020)}]{Fernandes20plb}%
  \BibitemOpen
  \bibfield  {author} {\bibinfo {author} {\bibfnamefont {P.~G.~S.}\
  \bibnamefont {{Fernandes}}},\ }\bibfield  {title} {\bibinfo {title} {{Charged
  black holes in AdS spaces in 4D Einstein Gauss-Bonnet gravity}},\ }\href
  {https://doi.org/10.1016/j.physletb.2020.135468} {\bibfield  {journal}
  {\bibinfo  {journal} {Phys. Lett. B}\ }\textbf {\bibinfo {volume} {805}},\
  \bibinfo {eid} {135468} (\bibinfo {year} {2020})},\ \Eprint
  {https://arxiv.org/abs/2003.05491} {arXiv:2003.05491 [gr-qc]} \BibitemShut
  {NoStop}%
\bibitem [{\citenamefont {{Hennigar}}\ \emph
  {et~al.}(2020{\natexlab{b}})\citenamefont {{Hennigar}}, \citenamefont
  {{Kubiz{\v{n}}{\'a}k}}, \citenamefont {{Mann}},\ and\ \citenamefont
  {{Pollack}}}]{Hennigar20PLB}%
  \BibitemOpen
  \bibfield  {author} {\bibinfo {author} {\bibfnamefont {R.~A.}\ \bibnamefont
  {{Hennigar}}}, \bibinfo {author} {\bibfnamefont {D.}~\bibnamefont
  {{Kubiz{\v{n}}{\'a}k}}}, \bibinfo {author} {\bibfnamefont {R.~B.}\
  \bibnamefont {{Mann}}},\ and\ \bibinfo {author} {\bibfnamefont
  {C.}~\bibnamefont {{Pollack}}},\ }\bibfield  {title} {\bibinfo {title}
  {{Lower-dimensional Gauss-Bonnet gravity and BTZ black holes}},\ }\href
  {https://doi.org/10.1016/j.physletb.2020.135657} {\bibfield  {journal}
  {\bibinfo  {journal} {Phys. Lett. B}\ }\textbf {\bibinfo {volume} {808}},\
  \bibinfo {eid} {135657} (\bibinfo {year} {2020}{\natexlab{b}})},\ \Eprint
  {https://arxiv.org/abs/2004.12995} {arXiv:2004.12995 [gr-qc]} \BibitemShut
  {NoStop}%
\bibitem [{\citenamefont {{Shaymatov}}\ and\ \citenamefont
  {{Dadhich}}(2022)}]{Shaymatov22JCAP}%
  \BibitemOpen
  \bibfield  {author} {\bibinfo {author} {\bibfnamefont {S.}~\bibnamefont
  {{Shaymatov}}}\ and\ \bibinfo {author} {\bibfnamefont {N.}~\bibnamefont
  {{Dadhich}}},\ }\bibfield  {title} {\bibinfo {title} {{Weak cosmic censorship
  conjecture in the pure Lovelock gravity}},\ }\href
  {https://doi.org/10.1088/1475-7516/2022/10/060} {\bibfield  {journal}
  {\bibinfo  {journal} {JCAP}\ }\textbf {\bibinfo {volume} {2022}}\bibfield
  {number} {\bibinfo  {number} { (10)},\ \bibinfo {eid} {060}},\ }\Eprint
  {https://arxiv.org/abs/2008.04092} {arXiv:2008.04092 [gr-qc]} \BibitemShut
  {NoStop}%
\bibitem [{\citenamefont {{Dadhich}}\ and\ \citenamefont
  {{Shaymatov}}(2022)}]{Dadhich22a}%
  \BibitemOpen
  \bibfield  {author} {\bibinfo {author} {\bibfnamefont {N.}~\bibnamefont
  {{Dadhich}}}\ and\ \bibinfo {author} {\bibfnamefont {S.}~\bibnamefont
  {{Shaymatov}}},\ }\bibfield  {title} {\bibinfo {title} {{Circular orbits
  around higher dimensional Einstein and pure Gauss-Bonnet rotating black
  holes}},\ }\href {https://doi.org/10.1016/j.dark.2022.100986} {\bibfield
  {journal} {\bibinfo  {journal} {Phys. Dark Universe}\ }\textbf {\bibinfo
  {volume} {35}},\ \bibinfo {eid} {100986} (\bibinfo {year} {2022})},\ \Eprint
  {https://arxiv.org/abs/2104.00427} {arXiv:2104.00427 [gr-qc]} \BibitemShut
  {NoStop}%
\bibitem [{\citenamefont {{Wu}}\ \emph {et~al.}(2021)\citenamefont {{Wu}},
  \citenamefont {{Hu}},\ and\ \citenamefont {{Xu}}}]{Wu21egb}%
  \BibitemOpen
  \bibfield  {author} {\bibinfo {author} {\bibfnamefont {C.-H.}\ \bibnamefont
  {{Wu}}}, \bibinfo {author} {\bibfnamefont {Y.-P.}\ \bibnamefont {{Hu}}},\
  and\ \bibinfo {author} {\bibfnamefont {H.}~\bibnamefont {{Xu}}},\ }\bibfield
  {title} {\bibinfo {title} {{Hawking evaporation of Einstein-Gauss-Bonnet AdS
  black holes in D {\ensuremath{\geqslant}}4 dimensions}},\ }\href
  {https://doi.org/10.1140/epjc/s10052-021-09140-6} {\bibfield  {journal}
  {\bibinfo  {journal} {Eur. Phys. J. C}\ }\textbf {\bibinfo {volume} {81}},\
  \bibinfo {eid} {351} (\bibinfo {year} {2021})},\ \Eprint
  {https://arxiv.org/abs/2103.00257} {arXiv:2103.00257 [hep-th]} \BibitemShut
  {NoStop}%
\bibitem [{\citenamefont {{Penrose}}(1969)}]{Penrose:1969pc}%
  \BibitemOpen
  \bibfield  {author} {\bibinfo {author} {\bibfnamefont {R.}~\bibnamefont
  {{Penrose}}},\ }\bibfield  {title} {\bibinfo {title} {{Gravitational
  Collapse: the Role of General Relativity}},\ }\href@noop {} {\bibfield
  {journal} {\bibinfo  {journal} {Riv. Nuovo Cim.}\ }\textbf {\bibinfo {volume}
  {1}},\ \bibinfo {pages} {252} (\bibinfo {year} {1969})}\BibitemShut {NoStop}%
\bibitem [{\citenamefont {{Abdujabbarov}}\ \emph {et~al.}(2011)\citenamefont
  {{Abdujabbarov}}, \citenamefont {{Ahmedov}}, \citenamefont {{Shaymatov}},\
  and\ \citenamefont {{Rakhmatov}}}]{Abdujabbarov11}%
  \BibitemOpen
  \bibfield  {author} {\bibinfo {author} {\bibfnamefont {A.~A.}\ \bibnamefont
  {{Abdujabbarov}}}, \bibinfo {author} {\bibfnamefont {B.~J.}\ \bibnamefont
  {{Ahmedov}}}, \bibinfo {author} {\bibfnamefont {S.~R.}\ \bibnamefont
  {{Shaymatov}}},\ and\ \bibinfo {author} {\bibfnamefont {A.~S.}\ \bibnamefont
  {{Rakhmatov}}},\ }\bibfield  {title} {\bibinfo {title} {{Penrose process in
  Kerr-Taub-NUT spacetime}},\ }\href
  {https://doi.org/10.1007/s10509-011-0740-8} {\bibfield  {journal} {\bibinfo
  {journal} {Astrophys Space Sci}\ }\textbf {\bibinfo {volume} {334}},\
  \bibinfo {pages} {237} (\bibinfo {year} {2011})},\ \Eprint
  {https://arxiv.org/abs/1105.1910} {arXiv:1105.1910 [astro-ph.SR]}
  \BibitemShut {NoStop}%
\bibitem [{\citenamefont {Toshmatov}\ \emph {et~al.}(2015)\citenamefont
  {Toshmatov}, \citenamefont {Abdujabbarov}, \citenamefont {Ahmedov},\ and\
  \citenamefont {Stuchl\'\i{}k}}]{Toshmatov:2014qja}%
  \BibitemOpen
  \bibfield  {author} {\bibinfo {author} {\bibfnamefont {B.}~\bibnamefont
  {Toshmatov}}, \bibinfo {author} {\bibfnamefont {A.}~\bibnamefont
  {Abdujabbarov}}, \bibinfo {author} {\bibfnamefont {B.}~\bibnamefont
  {Ahmedov}},\ and\ \bibinfo {author} {\bibfnamefont {Z.}~\bibnamefont
  {Stuchl\'\i{}k}},\ }\bibfield  {title} {\bibinfo {title} {{Particle motion
  and Penrose processes around rotating regular black hole}},\ }\href
  {https://doi.org/10.1007/s10509-015-2289-4} {\bibfield  {journal} {\bibinfo
  {journal} {Astrophys. Space Sci.}\ }\textbf {\bibinfo {volume} {357}},\
  \bibinfo {pages} {41} (\bibinfo {year} {2015})},\ \Eprint
  {https://arxiv.org/abs/1407.3697} {arXiv:1407.3697 [gr-qc]} \BibitemShut
  {NoStop}%
\bibitem [{\citenamefont {{Okabayashi}}\ and\ \citenamefont
  {{Maeda}}(2020)}]{Okabayashi20}%
  \BibitemOpen
  \bibfield  {author} {\bibinfo {author} {\bibfnamefont {K.}~\bibnamefont
  {{Okabayashi}}}\ and\ \bibinfo {author} {\bibfnamefont {K.-i.}\ \bibnamefont
  {{Maeda}}},\ }\bibfield  {title} {\bibinfo {title} {{Maximal efficiency of
  the collisional Penrose process with a spinning particle. II. Collision with
  a particle on the innermost stable circular orbit}},\ }\href
  {https://doi.org/10.1093/ptep/ptz143} {\bibfield  {journal} {\bibinfo
  {journal} {Prog. Theor. Exp. Phys.}\ }\textbf {\bibinfo {volume} {2020}},\
  \bibinfo {eid} {013E01} (\bibinfo {year} {2020})},\ \Eprint
  {https://arxiv.org/abs/1907.07126} {arXiv:1907.07126 [gr-qc]} \BibitemShut
  {NoStop}%
\bibitem [{\citenamefont {{Prabhu}}\ and\ \citenamefont
  {{Dadhich}}(2010)}]{Prabhu10}%
  \BibitemOpen
  \bibfield  {author} {\bibinfo {author} {\bibfnamefont {K.}~\bibnamefont
  {{Prabhu}}}\ and\ \bibinfo {author} {\bibfnamefont {N.}~\bibnamefont
  {{Dadhich}}},\ }\bibfield  {title} {\bibinfo {title} {{Energetics of a
  rotating charged black hole in 5-dimensional supergravity}},\ }\href
  {https://doi.org/10.1103/PhysRevD.81.024011} {\bibfield  {journal} {\bibinfo
  {journal} {Phys. Rev. D}\ }\textbf {\bibinfo {volume} {81}},\ \bibinfo {eid}
  {024011} (\bibinfo {year} {2010})},\ \Eprint
  {https://arxiv.org/abs/0902.3079} {arXiv:0902.3079 [hep-th]} \BibitemShut
  {NoStop}%
\bibitem [{\citenamefont {{Nozawa}}\ and\ \citenamefont
  {{Maeda}}(2005)}]{Nozawa05}%
  \BibitemOpen
  \bibfield  {author} {\bibinfo {author} {\bibfnamefont {M.}~\bibnamefont
  {{Nozawa}}}\ and\ \bibinfo {author} {\bibfnamefont {K.-I.}\ \bibnamefont
  {{Maeda}}},\ }\bibfield  {title} {\bibinfo {title} {{Energy extraction from
  higher dimensional black holes and black rings}},\ }\href
  {https://doi.org/10.1103/PhysRevD.71.084028} {\bibfield  {journal} {\bibinfo
  {journal} {Phys. Rev. D}\ }\textbf {\bibinfo {volume} {71}},\ \bibinfo {eid}
  {084028} (\bibinfo {year} {2005})},\ \Eprint
  {https://arxiv.org/abs/hep-th/0502166} {hep-th/0502166} \BibitemShut
  {NoStop}%
\bibitem [{\citenamefont {{Shaymatov}}(2024)}]{Shaymatov:2024MPP1}%
  \BibitemOpen
  \bibfield  {author} {\bibinfo {author} {\bibfnamefont {S.}~\bibnamefont
  {{Shaymatov}}},\ }\bibfield  {title} {\bibinfo {title} {{Efficiency of
  magnetic Penrose process in higher dimensional Myers-Perry black hole
  spacetimes}},\ }\href {https://doi.org/10.48550/arXiv.2402.02471} {\bibfield
  {journal} {\bibinfo  {journal} {arXiv e-prints}\ ,\ \bibinfo {eid}
  {arXiv:2402.02471}} (\bibinfo {year} {2024})},\ \Eprint
  {https://arxiv.org/abs/2402.02471} {arXiv:2402.02471 [gr-qc]} \BibitemShut
  {NoStop}%
\bibitem [{\citenamefont {{Shaymatov}}\ \emph {et~al.}(2024)\citenamefont
  {{Shaymatov}}, \citenamefont {{Dadhich}},\ and\ \citenamefont
  {{Tursunov}}}]{Shaymatov:2024MPP2}%
  \BibitemOpen
  \bibfield  {author} {\bibinfo {author} {\bibfnamefont {S.}~\bibnamefont
  {{Shaymatov}}}, \bibinfo {author} {\bibfnamefont {N.}~\bibnamefont
  {{Dadhich}}},\ and\ \bibinfo {author} {\bibfnamefont {A.}~\bibnamefont
  {{Tursunov}}},\ }\bibfield  {title} {\bibinfo {title} {{Energetics of
  Buchdahl stars and the magnetic Penrose process}},\ }\href
  {https://doi.org/10.48550/arXiv.2404.06870} {\bibfield  {journal} {\bibinfo
  {journal} {arXiv e-prints}\ ,\ \bibinfo {eid} {arXiv:2404.06870}} (\bibinfo
  {year} {2024})},\ \Eprint {https://arxiv.org/abs/2404.06870}
  {arXiv:2404.06870 [gr-qc]} \BibitemShut {NoStop}%
\bibitem [{\citenamefont {{Bardeen}}\ \emph {et~al.}(1972)\citenamefont
  {{Bardeen}}, \citenamefont {{Press}},\ and\ \citenamefont
  {{Teukolsky}}}]{Bardeen72}%
  \BibitemOpen
  \bibfield  {author} {\bibinfo {author} {\bibfnamefont {J.~M.}\ \bibnamefont
  {{Bardeen}}}, \bibinfo {author} {\bibfnamefont {W.~H.}\ \bibnamefont
  {{Press}}},\ and\ \bibinfo {author} {\bibfnamefont {S.~A.}\ \bibnamefont
  {{Teukolsky}}},\ }\bibfield  {title} {\bibinfo {title} {{Rotating Black
  Holes: Locally Nonrotating Frames, Energy Extraction, and Scalar Synchrotron
  Radiation}},\ }\href {https://doi.org/10.1086/151796} {\bibfield  {journal}
  {\bibinfo  {journal} {Astrophys. J.}\ }\textbf {\bibinfo {volume} {178}},\
  \bibinfo {pages} {347} (\bibinfo {year} {1972})}\BibitemShut {NoStop}%
\bibitem [{\citenamefont {{Wald}}(1974)}]{Wald74ApJ}%
  \BibitemOpen
  \bibfield  {author} {\bibinfo {author} {\bibfnamefont {R.~M.}\ \bibnamefont
  {{Wald}}},\ }\bibfield  {title} {\bibinfo {title} {{Energy Limits on the
  Penrose Process}},\ }\href {https://doi.org/10.1086/152959} {\bibfield
  {journal} {\bibinfo  {journal} {Astrophys. J.}\ }\textbf {\bibinfo {volume}
  {191}},\ \bibinfo {pages} {231} (\bibinfo {year} {1974})}\BibitemShut
  {NoStop}%
\bibitem [{\citenamefont {{Bhat}}\ \emph {et~al.}(1985)\citenamefont {{Bhat}},
  \citenamefont {{Dhurandhar}},\ and\ \citenamefont {{Dadhich}}}]{Bhat85}%
  \BibitemOpen
  \bibfield  {author} {\bibinfo {author} {\bibfnamefont {M.}~\bibnamefont
  {{Bhat}}}, \bibinfo {author} {\bibfnamefont {S.}~\bibnamefont
  {{Dhurandhar}}},\ and\ \bibinfo {author} {\bibfnamefont {N.}~\bibnamefont
  {{Dadhich}}},\ }\bibfield  {title} {\bibinfo {title} {{Energetics of the
  Kerr-Newman black hole by the Penrose process}},\ }\href
  {https://doi.org/10.1007/BF02715080} {\bibfield  {journal} {\bibinfo
  {journal} {Journal of Astrophysics and Astronomy}\ }\textbf {\bibinfo
  {volume} {6}},\ \bibinfo {pages} {85} (\bibinfo {year} {1985})}\BibitemShut
  {NoStop}%
\bibitem [{\citenamefont {{Parthasarathy}}\ \emph {et~al.}(1986)\citenamefont
  {{Parthasarathy}}, \citenamefont {{Wagh}}, \citenamefont {{Dhurandhar}},\
  and\ \citenamefont {{Dadhich}}}]{Parthasarathy86}%
  \BibitemOpen
  \bibfield  {author} {\bibinfo {author} {\bibfnamefont {S.}~\bibnamefont
  {{Parthasarathy}}}, \bibinfo {author} {\bibfnamefont {S.~M.}\ \bibnamefont
  {{Wagh}}}, \bibinfo {author} {\bibfnamefont {S.~V.}\ \bibnamefont
  {{Dhurandhar}}},\ and\ \bibinfo {author} {\bibfnamefont {N.}~\bibnamefont
  {{Dadhich}}},\ }\bibfield  {title} {\bibinfo {title} {{High Efficiency of the
  Penrose Process of Energy Extraction from Rotating Black Holes Immersed in
  Electromagnetic Fields}},\ }\href {https://doi.org/10.1086/164390} {\bibfield
   {journal} {\bibinfo  {journal} {Astrophys. J}\ }\textbf {\bibinfo {volume}
  {307}},\ \bibinfo {pages} {38} (\bibinfo {year} {1986})}\BibitemShut
  {NoStop}%
\bibitem [{\citenamefont {{Wagh}}\ and\ \citenamefont
  {{Dadhich}}(1989)}]{Wagh89}%
  \BibitemOpen
  \bibfield  {author} {\bibinfo {author} {\bibfnamefont {S.~M.}\ \bibnamefont
  {{Wagh}}}\ and\ \bibinfo {author} {\bibfnamefont {N.}~\bibnamefont
  {{Dadhich}}},\ }\bibfield  {title} {\bibinfo {title} {{The energetics of
  black holes in electromagnetic fields by the penrose process}},\ }\href
  {https://doi.org/10.1016/0370-1573(89)90156-7} {\bibfield  {journal}
  {\bibinfo  {journal} {Phys. Rep.}\ }\textbf {\bibinfo {volume} {183}},\
  \bibinfo {pages} {137} (\bibinfo {year} {1989})}\BibitemShut {NoStop}%
\bibitem [{\citenamefont {{Alic}}\ \emph {et~al.}(2012)\citenamefont {{Alic}},
  \citenamefont {{Moesta}}, \citenamefont {{Rezzolla}}, \citenamefont
  {{Zanotti}},\ and\ \citenamefont {{Jaramillo}}}]{Alic12ApJ}%
  \BibitemOpen
  \bibfield  {author} {\bibinfo {author} {\bibfnamefont {D.}~\bibnamefont
  {{Alic}}}, \bibinfo {author} {\bibfnamefont {P.}~\bibnamefont {{Moesta}}},
  \bibinfo {author} {\bibfnamefont {L.}~\bibnamefont {{Rezzolla}}}, \bibinfo
  {author} {\bibfnamefont {O.}~\bibnamefont {{Zanotti}}},\ and\ \bibinfo
  {author} {\bibfnamefont {J.~L.}\ \bibnamefont {{Jaramillo}}},\ }\bibfield
  {title} {\bibinfo {title} {{Accurate Simulations of Binary Black Hole Mergers
  in Force-free Electrodynamics}},\ }\href
  {https://doi.org/10.1088/0004-637X/754/1/36} {\bibfield  {journal} {\bibinfo
  {journal} {Astrophys. J.}\ }\textbf {\bibinfo {volume} {754}},\ \bibinfo
  {eid} {36} (\bibinfo {year} {2012})},\ \Eprint
  {https://arxiv.org/abs/1204.2226} {arXiv:1204.2226 [gr-qc]} \BibitemShut
  {NoStop}%
\bibitem [{\citenamefont {{Moesta}}\ \emph {et~al.}(2012)\citenamefont
  {{Moesta}}, \citenamefont {{Alic}}, \citenamefont {{Rezzolla}}, \citenamefont
  {{Zanotti}},\ and\ \citenamefont {{Palenzuela}}}]{Moesta12ApJ}%
  \BibitemOpen
  \bibfield  {author} {\bibinfo {author} {\bibfnamefont {P.}~\bibnamefont
  {{Moesta}}}, \bibinfo {author} {\bibfnamefont {D.}~\bibnamefont {{Alic}}},
  \bibinfo {author} {\bibfnamefont {L.}~\bibnamefont {{Rezzolla}}}, \bibinfo
  {author} {\bibfnamefont {O.}~\bibnamefont {{Zanotti}}},\ and\ \bibinfo
  {author} {\bibfnamefont {C.}~\bibnamefont {{Palenzuela}}},\ }\bibfield
  {title} {\bibinfo {title} {{On the Detectability of Dual Jets from Binary
  Black Holes}},\ }\href {https://doi.org/10.1088/2041-8205/749/2/L32}
  {\bibfield  {journal} {\bibinfo  {journal} {Astrophys. J.}\ }\textbf
  {\bibinfo {volume} {749}},\ \bibinfo {eid} {L32} (\bibinfo {year} {2012})},\
  \Eprint {https://arxiv.org/abs/1109.1177} {arXiv:1109.1177 [gr-qc]}
  \BibitemShut {NoStop}%
\bibitem [{\citenamefont {Dadhich}\ \emph {et~al.}(2018)\citenamefont
  {Dadhich}, \citenamefont {Tursunov}, \citenamefont {Ahmedov},\ and\
  \citenamefont {Stuchl\'\i{}k}}]{Dadhich18mnras}%
  \BibitemOpen
  \bibfield  {author} {\bibinfo {author} {\bibfnamefont {N.}~\bibnamefont
  {Dadhich}}, \bibinfo {author} {\bibfnamefont {A.}~\bibnamefont {Tursunov}},
  \bibinfo {author} {\bibfnamefont {B.}~\bibnamefont {Ahmedov}},\ and\ \bibinfo
  {author} {\bibfnamefont {Z.}~\bibnamefont {Stuchl\'\i{}k}},\ }\bibfield
  {title} {\bibinfo {title} {{The distinguishing signature of Magnetic Penrose
  Process}},\ }\href {https://doi.org/10.1093/mnrasl/sly073} {\bibfield
  {journal} {\bibinfo  {journal} {Mon. Not. Roy. Astron. Soc.}\ }\textbf
  {\bibinfo {volume} {478}},\ \bibinfo {pages} {L89} (\bibinfo {year}
  {2018})},\ \Eprint {https://arxiv.org/abs/1804.09679} {arXiv:1804.09679
  [astro-ph.HE]} \BibitemShut {NoStop}%
\bibitem [{\citenamefont {Tursunov}\ and\ \citenamefont
  {Dadhich}(2019)}]{Tursunov:2019oiq}%
  \BibitemOpen
  \bibfield  {author} {\bibinfo {author} {\bibfnamefont {A.}~\bibnamefont
  {Tursunov}}\ and\ \bibinfo {author} {\bibfnamefont {N.}~\bibnamefont
  {Dadhich}},\ }\bibfield  {title} {\bibinfo {title} {{Fifty years of energy
  extraction from rotating black hole: revisiting magnetic Penrose process}},\
  }\href {https://doi.org/10.3390/universe5050125} {\bibfield  {journal}
  {\bibinfo  {journal} {Universe}\ }\textbf {\bibinfo {volume} {5}},\ \bibinfo
  {pages} {125} (\bibinfo {year} {2019})},\ \Eprint
  {https://arxiv.org/abs/1905.05321} {arXiv:1905.05321 [astro-ph.HE]}
  \BibitemShut {NoStop}%
\bibitem [{\citenamefont {{Tursunov}}\ \emph {et~al.}(2020)\citenamefont
  {{Tursunov}}, \citenamefont {{Stuchl{\'\i}k}}, \citenamefont {{Kolo{\v{s}}}},
  \citenamefont {{Dadhich}},\ and\ \citenamefont {{Ahmedov}}}]{Tursunov20ApJ}%
  \BibitemOpen
  \bibfield  {author} {\bibinfo {author} {\bibfnamefont {A.}~\bibnamefont
  {{Tursunov}}}, \bibinfo {author} {\bibfnamefont {Z.}~\bibnamefont
  {{Stuchl{\'\i}k}}}, \bibinfo {author} {\bibfnamefont {M.}~\bibnamefont
  {{Kolo{\v{s}}}}}, \bibinfo {author} {\bibfnamefont {N.}~\bibnamefont
  {{Dadhich}}},\ and\ \bibinfo {author} {\bibfnamefont {B.}~\bibnamefont
  {{Ahmedov}}},\ }\bibfield  {title} {\bibinfo {title} {{Supermassive Black
  Holes as Possible Sources of Ultrahigh-energy Cosmic Rays}},\ }\href
  {https://doi.org/10.3847/1538-4357/ab8ae9} {\bibfield  {journal} {\bibinfo
  {journal} {Astrophys. J.}\ }\textbf {\bibinfo {volume} {895}},\ \bibinfo
  {eid} {14} (\bibinfo {year} {2020})}\BibitemShut {NoStop}%
\bibitem [{\citenamefont {{Shaymatov}}\ \emph {et~al.}(2022)\citenamefont
  {{Shaymatov}}, \citenamefont {{Sheoran}}, \citenamefont {{Becerril}},
  \citenamefont {{Nucamendi}},\ and\ \citenamefont {{Ahmedov}}}]{Shaymatov22b}%
  \BibitemOpen
  \bibfield  {author} {\bibinfo {author} {\bibfnamefont {S.}~\bibnamefont
  {{Shaymatov}}}, \bibinfo {author} {\bibfnamefont {P.}~\bibnamefont
  {{Sheoran}}}, \bibinfo {author} {\bibfnamefont {R.}~\bibnamefont
  {{Becerril}}}, \bibinfo {author} {\bibfnamefont {U.}~\bibnamefont
  {{Nucamendi}}},\ and\ \bibinfo {author} {\bibfnamefont {B.}~\bibnamefont
  {{Ahmedov}}},\ }\bibfield  {title} {\bibinfo {title} {{Efficiency of Penrose
  process in spacetime of axially symmetric magnetized Reissner-Nordstr{\"o}m
  black hole}},\ }\href {https://doi.org/10.1103/PhysRevD.106.024039}
  {\bibfield  {journal} {\bibinfo  {journal} {Phys. Rev. D}\ }\textbf {\bibinfo
  {volume} {106}},\ \bibinfo {eid} {024039} (\bibinfo {year}
  {2022})}\BibitemShut {NoStop}%
\bibitem [{\citenamefont {{Denardo}}\ and\ \citenamefont
  {{Ruffini}}(1973)}]{Denardo73PLB}%
  \BibitemOpen
  \bibfield  {author} {\bibinfo {author} {\bibfnamefont {G.}~\bibnamefont
  {{Denardo}}}\ and\ \bibinfo {author} {\bibfnamefont {R.}~\bibnamefont
  {{Ruffini}}},\ }\bibfield  {title} {\bibinfo {title} {{On the energetics of
  Reissner Nordstr{\o}m geometries}},\ }\href
  {https://doi.org/10.1016/0370-2693(73)90198-6} {\bibfield  {journal}
  {\bibinfo  {journal} {Phys. Lett. B}\ }\textbf {\bibinfo {volume} {45}},\
  \bibinfo {pages} {259} (\bibinfo {year} {1973})}\BibitemShut {NoStop}%
\bibitem [{\citenamefont {{Tursunov}}\ \emph {et~al.}(2021)\citenamefont
  {{Tursunov}}, \citenamefont {{Juraev}}, \citenamefont {{Stuchl{\'\i}k}},\
  and\ \citenamefont {{Kolo{\v{s}}}}}]{Tursunov21EPP}%
  \BibitemOpen
  \bibfield  {author} {\bibinfo {author} {\bibfnamefont {A.}~\bibnamefont
  {{Tursunov}}}, \bibinfo {author} {\bibfnamefont {B.}~\bibnamefont
  {{Juraev}}}, \bibinfo {author} {\bibfnamefont {Z.}~\bibnamefont
  {{Stuchl{\'\i}k}}},\ and\ \bibinfo {author} {\bibfnamefont {M.}~\bibnamefont
  {{Kolo{\v{s}}}}},\ }\bibfield  {title} {\bibinfo {title} {{Electric Penrose
  process: High-energy acceleration of ionized particles by nonrotating weakly
  charged black hole}},\ }\href {https://doi.org/10.1103/PhysRevD.104.084099}
  {\bibfield  {journal} {\bibinfo  {journal} {Phys. Rev. D}\ }\textbf {\bibinfo
  {volume} {104}},\ \bibinfo {eid} {084099} (\bibinfo {year} {2021})},\ \Eprint
  {https://arxiv.org/abs/2109.10288} {arXiv:2109.10288 [gr-qc]} \BibitemShut
  {NoStop}%
\bibitem [{\citenamefont {Stuchlík}\ \emph {et~al.}(2021)\citenamefont
  {Stuchlík}, \citenamefont {Kološ},\ and\ \citenamefont
  {Tursunov}}]{Stuchlik:2021}%
  \BibitemOpen
  \bibfield  {author} {\bibinfo {author} {\bibfnamefont {Z.}~\bibnamefont
  {Stuchlík}}, \bibinfo {author} {\bibfnamefont {M.}~\bibnamefont {Kološ}},\
  and\ \bibinfo {author} {\bibfnamefont {A.}~\bibnamefont {Tursunov}},\
  }\bibfield  {title} {\bibinfo {title} {Penrose process: Its variants and
  astrophysical applications},\ }\bibfield  {journal} {\bibinfo  {journal}
  {Universe}\ }\textbf {\bibinfo {volume} {7}},\ \href
  {https://doi.org/10.3390/universe7110416} {10.3390/universe7110416} (\bibinfo
  {year} {2021})\BibitemShut {NoStop}%
\bibitem [{\citenamefont {Alloqulov}\ \emph {et~al.}(2023)\citenamefont
  {Alloqulov}, \citenamefont {Narzilloev}, \citenamefont {Hussain},
  \citenamefont {Abdujabbarov},\ and\ \citenamefont
  {Ahmedov}}]{ALLOQULOV2023302}%
  \BibitemOpen
  \bibfield  {author} {\bibinfo {author} {\bibfnamefont {M.}~\bibnamefont
  {Alloqulov}}, \bibinfo {author} {\bibfnamefont {B.}~\bibnamefont
  {Narzilloev}}, \bibinfo {author} {\bibfnamefont {I.}~\bibnamefont {Hussain}},
  \bibinfo {author} {\bibfnamefont {A.}~\bibnamefont {Abdujabbarov}},\ and\
  \bibinfo {author} {\bibfnamefont {B.}~\bibnamefont {Ahmedov}},\ }\bibfield
  {title} {\bibinfo {title} {Energetic processes around electromagnetically
  charged black hole in the rastall gravity},\ }\href
  {https://doi.org/https://doi.org/10.1016/j.cjph.2023.07.005} {\bibfield
  {journal} {\bibinfo  {journal} {Chinese Journal of Physics}\ }\textbf
  {\bibinfo {volume} {85}},\ \bibinfo {pages} {302} (\bibinfo {year}
  {2023})}\BibitemShut {NoStop}%
\bibitem [{\citenamefont {{Baez}}\ \emph {et~al.}(2024)\citenamefont {{Baez}},
  \citenamefont {{Breton}},\ and\ \citenamefont
  {{Cabrera-Munguia}}}]{Baez2024epp}%
  \BibitemOpen
  \bibfield  {author} {\bibinfo {author} {\bibfnamefont {A.}~\bibnamefont
  {{Baez}}}, \bibinfo {author} {\bibfnamefont {N.}~\bibnamefont {{Breton}}},\
  and\ \bibinfo {author} {\bibfnamefont {I.}~\bibnamefont
  {{Cabrera-Munguia}}},\ }\bibfield  {title} {\bibinfo {title} {{Energy
  extraction from the Reissner-Nordstr{\"o}m de Sitter black hole}},\ }\href
  {https://doi.org/10.48550/arXiv.2406.12088} {\bibfield  {journal} {\bibinfo
  {journal} {arXiv e-prints}\ ,\ \bibinfo {eid} {arXiv:2406.12088}} (\bibinfo
  {year} {2024})},\ \Eprint {https://arxiv.org/abs/2406.12088}
  {arXiv:2406.12088 [gr-qc]} \BibitemShut {NoStop}%
\bibitem [{\citenamefont {{Fender}}\ \emph {et~al.}(2004)\citenamefont
  {{Fender}}, \citenamefont {{Belloni}},\ and\ \citenamefont
  {{Gallo}}}]{Fender04mnrs}%
  \BibitemOpen
  \bibfield  {author} {\bibinfo {author} {\bibfnamefont {R.~P.}\ \bibnamefont
  {{Fender}}}, \bibinfo {author} {\bibfnamefont {T.~M.}\ \bibnamefont
  {{Belloni}}},\ and\ \bibinfo {author} {\bibfnamefont {E.}~\bibnamefont
  {{Gallo}}},\ }\bibfield  {title} {\bibinfo {title} {{Towards a unified model
  for black hole X-ray binary jets}},\ }\href
  {https://doi.org/10.1111/j.1365-2966.2004.08384.x} {\bibfield  {journal}
  {\bibinfo  {journal} {Mon. Not. R. Astron. Soc.}\ }\textbf {\bibinfo {volume}
  {355}},\ \bibinfo {pages} {1105} (\bibinfo {year} {2004})},\ \Eprint
  {https://arxiv.org/abs/astro-ph/0409360} {arXiv:astro-ph/0409360 [astro-ph]}
  \BibitemShut {NoStop}%
\bibitem [{\citenamefont {{Auchettl}}\ \emph {et~al.}(2017)\citenamefont
  {{Auchettl}}, \citenamefont {{Guillochon}},\ and\ \citenamefont
  {{Ramirez-Ruiz}}}]{Auchettl17ApJ}%
  \BibitemOpen
  \bibfield  {author} {\bibinfo {author} {\bibfnamefont {K.}~\bibnamefont
  {{Auchettl}}}, \bibinfo {author} {\bibfnamefont {J.}~\bibnamefont
  {{Guillochon}}},\ and\ \bibinfo {author} {\bibfnamefont {E.}~\bibnamefont
  {{Ramirez-Ruiz}}},\ }\bibfield  {title} {\bibinfo {title} {{New Physical
  Insights about Tidal Disruption Events from a Comprehensive Observational
  Inventory at X-Ray Wavelengths}},\ }\href
  {https://doi.org/10.3847/1538-4357/aa633b} {\bibfield  {journal} {\bibinfo
  {journal} {Astrophys. J.}\ }\textbf {\bibinfo {volume} {838}},\ \bibinfo
  {eid} {149} (\bibinfo {year} {2017})},\ \Eprint
  {https://arxiv.org/abs/1611.02291} {arXiv:1611.02291 [astro-ph.HE]}
  \BibitemShut {NoStop}%
\bibitem [{\citenamefont {{The IceCube Collaboration}}\ and\ \citenamefont
  {et~al.}(2018)}]{IceCube17b}%
  \BibitemOpen
  \bibfield  {author} {\bibinfo {author} {\bibnamefont {{The IceCube
  Collaboration}}}\ and\ \bibinfo {author} {\bibnamefont {et~al.}},\ }\bibfield
   {title} {\bibinfo {title} {{Multimessenger observations of a flaring blazar
  coincident with high-energy neutrino IceCube-170922A}},\ }\href
  {https://doi.org/10.1126/science.aat1378} {\bibfield  {journal} {\bibinfo
  {journal} {Science}\ }\textbf {\bibinfo {volume} {361}},\ \bibinfo {eid}
  {eaat1378} (\bibinfo {year} {2018})},\ \Eprint
  {https://arxiv.org/abs/1807.08816} {arXiv:1807.08816 [gr-qc]} \BibitemShut
  {NoStop}%
\bibitem [{\citenamefont {{Abramowicz}}\ and\ \citenamefont
  {{Fragile}}(2013)}]{Abramowicz13}%
  \BibitemOpen
  \bibfield  {author} {\bibinfo {author} {\bibfnamefont {M.~A.}\ \bibnamefont
  {{Abramowicz}}}\ and\ \bibinfo {author} {\bibfnamefont {P.~C.}\ \bibnamefont
  {{Fragile}}},\ }\bibfield  {title} {\bibinfo {title} {{Foundations of Black
  Hole Accretion Disk Theory}},\ }\href {https://doi.org/10.12942/lrr-2013-1}
  {\bibfield  {journal} {\bibinfo  {journal} {Living Rev. Relativ.}\ }\textbf
  {\bibinfo {volume} {16}},\ \bibinfo {eid} {1} (\bibinfo {year} {2013})},\
  \Eprint {https://arxiv.org/abs/1104.5499} {arXiv:1104.5499 [astro-ph.HE]}
  \BibitemShut {NoStop}%
\bibitem [{\citenamefont {Tursunov}\ \emph {et~al.}(2021)\citenamefont
  {Tursunov}, \citenamefont {Juraev}, \citenamefont {Stuchl\'\i{}k},\ and\
  \citenamefont {Kolo\v{s}}}]{Tursunov:2021}%
  \BibitemOpen
  \bibfield  {author} {\bibinfo {author} {\bibfnamefont {A.}~\bibnamefont
  {Tursunov}}, \bibinfo {author} {\bibfnamefont {B.}~\bibnamefont {Juraev}},
  \bibinfo {author} {\bibfnamefont {Z.}~\bibnamefont {Stuchl\'\i{}k}},\ and\
  \bibinfo {author} {\bibfnamefont {M.}~\bibnamefont {Kolo\v{s}}},\ }\bibfield
  {title} {\bibinfo {title} {{Electric Penrose process: High-energy
  acceleration of ionized particles by nonrotating weakly charged black
  hole}},\ }\href {https://doi.org/10.1103/PhysRevD.104.084099} {\bibfield
  {journal} {\bibinfo  {journal} {Phys. Rev. D}\ }\textbf {\bibinfo {volume}
  {104}},\ \bibinfo {pages} {084099} (\bibinfo {year} {2021})},\ \Eprint
  {https://arxiv.org/abs/2109.10288} {arXiv:2109.10288 [gr-qc]} \BibitemShut
  {NoStop}%
\bibitem [{\citenamefont {Kurbonov}\ \emph {et~al.}(2023)\citenamefont
  {Kurbonov}, \citenamefont {Rayimbaev}, \citenamefont {Alloqulov},
  \citenamefont {Zahid}, \citenamefont {Abdulxamidov}, \citenamefont
  {Abdujabbarov},\ and\ \citenamefont {Kurbanova}}]{Kurbonov2023}%
  \BibitemOpen
  \bibfield  {author} {\bibinfo {author} {\bibfnamefont {N.}~\bibnamefont
  {Kurbonov}}, \bibinfo {author} {\bibfnamefont {J.}~\bibnamefont {Rayimbaev}},
  \bibinfo {author} {\bibfnamefont {M.}~\bibnamefont {Alloqulov}}, \bibinfo
  {author} {\bibfnamefont {M.}~\bibnamefont {Zahid}}, \bibinfo {author}
  {\bibfnamefont {F.}~\bibnamefont {Abdulxamidov}}, \bibinfo {author}
  {\bibfnamefont {A.}~\bibnamefont {Abdujabbarov}},\ and\ \bibinfo {author}
  {\bibfnamefont {M.}~\bibnamefont {Kurbanova}},\ }\bibfield  {title} {\bibinfo
  {title} {Charged particles and penrose process near charged black holes in
  einstein--maxwell-scalar theory},\ }\href
  {https://doi.org/10.1140/epjc/s10052-023-11691-9} {\bibfield  {journal}
  {\bibinfo  {journal} {The European Physical Journal C}\ }\textbf {\bibinfo
  {volume} {83}},\ \bibinfo {pages} {506} (\bibinfo {year} {2023})}\BibitemShut
  {NoStop}%
\bibitem [{\citenamefont {Bokhari}\ \emph {et~al.}(2020)\citenamefont
  {Bokhari}, \citenamefont {Rayimbaev},\ and\ \citenamefont
  {Ahmedov}}]{Bokhari20}%
  \BibitemOpen
  \bibfield  {author} {\bibinfo {author} {\bibfnamefont {A.~H.}\ \bibnamefont
  {Bokhari}}, \bibinfo {author} {\bibfnamefont {J.}~\bibnamefont {Rayimbaev}},\
  and\ \bibinfo {author} {\bibfnamefont {B.}~\bibnamefont {Ahmedov}},\
  }\bibfield  {title} {\bibinfo {title} {Test particles dynamics around
  deformed reissner-nordstr\"om black hole},\ }\href
  {https://doi.org/10.1103/PhysRevD.102.124078} {\bibfield  {journal} {\bibinfo
   {journal} {Phys. Rev. D}\ }\textbf {\bibinfo {volume} {102}},\ \bibinfo
  {pages} {124078} (\bibinfo {year} {2020})}\BibitemShut {NoStop}%
\bibitem [{\citenamefont {{Rayimbaev}}\ \emph {et~al.}(2021)\citenamefont
  {{Rayimbaev}}, \citenamefont {{Shaymatov}},\ and\ \citenamefont
  {{Jamil}}}]{Rayimbaev-Shaymatov21a}%
  \BibitemOpen
  \bibfield  {author} {\bibinfo {author} {\bibfnamefont {J.}~\bibnamefont
  {{Rayimbaev}}}, \bibinfo {author} {\bibfnamefont {S.}~\bibnamefont
  {{Shaymatov}}},\ and\ \bibinfo {author} {\bibfnamefont {M.}~\bibnamefont
  {{Jamil}}},\ }\bibfield  {title} {\bibinfo {title} {{Dynamics and epicyclic
  motions of particles around the Schwarzschild-de Sitter black hole in perfect
  fluid dark matter}},\ }\href
  {https://doi.org/10.1140/epjc/s10052-021-09488-9} {\bibfield  {journal}
  {\bibinfo  {journal} {Eur. Phys. J. C}\ }\textbf {\bibinfo {volume} {81}},\
  \bibinfo {eid} {699} (\bibinfo {year} {2021})},\ \Eprint
  {https://arxiv.org/abs/2107.13436} {arXiv:2107.13436 [gr-qc]} \BibitemShut
  {NoStop}%
\bibitem [{\citenamefont {Novikov}\ and\ \citenamefont
  {Thorne}(1973)}]{Novikov1973}%
  \BibitemOpen
  \bibfield  {author} {\bibinfo {author} {\bibfnamefont {I.~D.}\ \bibnamefont
  {Novikov}}\ and\ \bibinfo {author} {\bibfnamefont {K.~S.}\ \bibnamefont
  {Thorne}},\ }\href
  {http://inis.iaea.org/search/search.aspx?orig_q=RN:06167207} {\emph {\bibinfo
  {title} {Astrophysics of black holes}}}\ (\bibinfo  {publisher} {Gordon and
  Breach, Science Publishers, Inc},\ \bibinfo {address} {United States},\
  \bibinfo {year} {1973})\BibitemShut {NoStop}%
\bibitem [{\citenamefont {{Page}}\ and\ \citenamefont
  {{Thorne}}(1974)}]{1974ApJ...191..499P}%
  \BibitemOpen
  \bibfield  {author} {\bibinfo {author} {\bibfnamefont {D.~N.}\ \bibnamefont
  {{Page}}}\ and\ \bibinfo {author} {\bibfnamefont {K.~S.}\ \bibnamefont
  {{Thorne}}},\ }\bibfield  {title} {\bibinfo {title} {{Disk-Accretion onto a
  Black Hole. Time-Averaged Structure of Accretion Disk}},\ }\href
  {https://doi.org/10.1086/152990} {\bibfield  {journal} {\bibinfo  {journal}
  {\apj}\ }\textbf {\bibinfo {volume} {191}},\ \bibinfo {pages} {499} (\bibinfo
  {year} {1974})}\BibitemShut {NoStop}%
\bibitem [{\citenamefont {Kurmanov}\ \emph {et~al.}(2022)\citenamefont
  {Kurmanov}, \citenamefont {Boshkayev}, \citenamefont {Giambò}, \citenamefont
  {Konysbayev}, \citenamefont {Luongo}, \citenamefont {Malafarina},\ and\
  \citenamefont {Quevedo}}]{Kurmanov_2022}%
  \BibitemOpen
  \bibfield  {author} {\bibinfo {author} {\bibfnamefont {E.}~\bibnamefont
  {Kurmanov}}, \bibinfo {author} {\bibfnamefont {K.}~\bibnamefont {Boshkayev}},
  \bibinfo {author} {\bibfnamefont {R.}~\bibnamefont {Giambò}}, \bibinfo
  {author} {\bibfnamefont {T.}~\bibnamefont {Konysbayev}}, \bibinfo {author}
  {\bibfnamefont {O.}~\bibnamefont {Luongo}}, \bibinfo {author} {\bibfnamefont
  {D.}~\bibnamefont {Malafarina}},\ and\ \bibinfo {author} {\bibfnamefont
  {H.}~\bibnamefont {Quevedo}},\ }\bibfield  {title} {\bibinfo {title}
  {Accretion disk luminosity for black holes surrounded by dark matter with
  anisotropic pressure},\ }\href {https://doi.org/10.3847/1538-4357/ac41d4}
  {\bibfield  {journal} {\bibinfo  {journal} {The Astrophysical Journal}\
  }\textbf {\bibinfo {volume} {925}},\ \bibinfo {pages} {210} (\bibinfo {year}
  {2022})}\BibitemShut {NoStop}%
\bibitem [{\citenamefont {Shakura}\ and\ \citenamefont
  {Sunyaev}(1973)}]{Shakura:1972te}%
  \BibitemOpen
  \bibfield  {author} {\bibinfo {author} {\bibfnamefont {N.~I.}\ \bibnamefont
  {Shakura}}\ and\ \bibinfo {author} {\bibfnamefont {R.~A.}\ \bibnamefont
  {Sunyaev}},\ }\bibfield  {title} {\bibinfo {title} {{Black holes in binary
  systems. Observational appearance}},\ }\href@noop {} {\bibfield  {journal}
  {\bibinfo  {journal} {Astron. Astrophys.}\ }\textbf {\bibinfo {volume}
  {24}},\ \bibinfo {pages} {337} (\bibinfo {year} {1973})}\BibitemShut
  {NoStop}%
\bibitem [{\citenamefont {Thorne}(1974)}]{Thorne:1974ve}%
  \BibitemOpen
  \bibfield  {author} {\bibinfo {author} {\bibfnamefont {K.~S.}\ \bibnamefont
  {Thorne}},\ }\bibfield  {title} {\bibinfo {title} {{Disk accretion onto a
  black hole. 2. Evolution of the hole.}},\ }\href
  {https://doi.org/10.1086/152991} {\bibfield  {journal} {\bibinfo  {journal}
  {Astrophys. J.}\ }\textbf {\bibinfo {volume} {191}},\ \bibinfo {pages} {507}
  (\bibinfo {year} {1974})}\BibitemShut {NoStop}%
\bibitem [{\citenamefont {Shaymatov}\ \emph {et~al.}(2022)\citenamefont
  {Shaymatov}, \citenamefont {Jamil}, \citenamefont {Jusufi},\ and\
  \citenamefont {Bamba}}]{Shaymatov2022}%
  \BibitemOpen
  \bibfield  {author} {\bibinfo {author} {\bibfnamefont {S.}~\bibnamefont
  {Shaymatov}}, \bibinfo {author} {\bibfnamefont {M.}~\bibnamefont {Jamil}},
  \bibinfo {author} {\bibfnamefont {K.}~\bibnamefont {Jusufi}},\ and\ \bibinfo
  {author} {\bibfnamefont {K.}~\bibnamefont {Bamba}},\ }\bibfield  {title}
  {\bibinfo {title} {Constraints on the magnetized ernst black hole spacetime
  through quasiperiodic oscillations},\ }\href
  {https://doi.org/10.1140/epjc/s10052-022-10560-1} {\bibfield  {journal}
  {\bibinfo  {journal} {The European Physical Journal C}\ }\textbf {\bibinfo
  {volume} {82}},\ \bibinfo {pages} {636} (\bibinfo {year} {2022})}\BibitemShut
  {NoStop}%
\bibitem [{\citenamefont {{Shaymatov}}\ \emph {et~al.}(2023)\citenamefont
  {{Shaymatov}}, \citenamefont {{Jusufi}}, \citenamefont {{Alloqulov}},\ and\
  \citenamefont {{Ahmedov}}}]{Shaymatov23EPJP}%
  \BibitemOpen
  \bibfield  {author} {\bibinfo {author} {\bibfnamefont {S.}~\bibnamefont
  {{Shaymatov}}}, \bibinfo {author} {\bibfnamefont {K.}~\bibnamefont
  {{Jusufi}}}, \bibinfo {author} {\bibfnamefont {M.}~\bibnamefont
  {{Alloqulov}}},\ and\ \bibinfo {author} {\bibfnamefont {B.}~\bibnamefont
  {{Ahmedov}}},\ }\bibfield  {title} {\bibinfo {title} {{Epicyclic motions and
  constraints on the charged stringy black hole spacetime}},\ }\href
  {https://doi.org/10.1140/epjp/s13360-023-04604-y} {\bibfield  {journal}
  {\bibinfo  {journal} {Eur. Phys. J. Plus}\ }\textbf {\bibinfo {volume}
  {138}},\ \bibinfo {eid} {997} (\bibinfo {year} {2023})},\ \Eprint
  {https://arxiv.org/abs/2303.06494} {arXiv:2303.06494 [gr-qc]} \BibitemShut
  {NoStop}%
\bibitem [{\citenamefont {Boshkayev}\ \emph {et~al.}(2020)\citenamefont
  {Boshkayev}, \citenamefont {Idrissov}, \citenamefont {Luongo},\ and\
  \citenamefont {Malafarina}}]{Boshkayev:2020kle}%
  \BibitemOpen
  \bibfield  {author} {\bibinfo {author} {\bibfnamefont {K.}~\bibnamefont
  {Boshkayev}}, \bibinfo {author} {\bibfnamefont {A.}~\bibnamefont {Idrissov}},
  \bibinfo {author} {\bibfnamefont {O.}~\bibnamefont {Luongo}},\ and\ \bibinfo
  {author} {\bibfnamefont {D.}~\bibnamefont {Malafarina}},\ }\bibfield  {title}
  {\bibinfo {title} {{Accretion disc luminosity for black holes surrounded by
  dark matter}},\ }\href {https://doi.org/10.1093/mnras/staa1564} {\bibfield
  {journal} {\bibinfo  {journal} {Mon. Not. Roy. Astron. Soc.}\ }\textbf
  {\bibinfo {volume} {496}},\ \bibinfo {pages} {1115} (\bibinfo {year}
  {2020})},\ \Eprint {https://arxiv.org/abs/2006.01269} {arXiv:2006.01269
  [astro-ph.HE]} \BibitemShut {NoStop}%
\bibitem [{\citenamefont {Narzilloev}\ and\ \citenamefont
  {Ahmedov}(2022)}]{sym14091765}%
  \BibitemOpen
  \bibfield  {author} {\bibinfo {author} {\bibfnamefont {B.}~\bibnamefont
  {Narzilloev}}\ and\ \bibinfo {author} {\bibfnamefont {B.}~\bibnamefont
  {Ahmedov}},\ }\bibfield  {title} {\bibinfo {title} {Radiation properties of
  the accretion disk around a black hole surrounded by pfdm},\ }\bibfield
  {journal} {\bibinfo  {journal} {Symmetry}\ }\textbf {\bibinfo {volume}
  {14}},\ \href {https://doi.org/10.3390/sym14091765} {10.3390/sym14091765}
  (\bibinfo {year} {2022})\BibitemShut {NoStop}%
\bibitem [{\citenamefont {{Alloqulov}}\ \emph {et~al.}(2024)\citenamefont
  {{Alloqulov}}, \citenamefont {{Shaymatov}}, \citenamefont {{Ahmedov}},\ and\
  \citenamefont {{Jawad}}}]{Alloqulov24CPC}%
  \BibitemOpen
  \bibfield  {author} {\bibinfo {author} {\bibfnamefont {M.}~\bibnamefont
  {{Alloqulov}}}, \bibinfo {author} {\bibfnamefont {S.}~\bibnamefont
  {{Shaymatov}}}, \bibinfo {author} {\bibfnamefont {B.}~\bibnamefont
  {{Ahmedov}}},\ and\ \bibinfo {author} {\bibfnamefont {A.}~\bibnamefont
  {{Jawad}}},\ }\bibfield  {title} {\bibinfo {title} {{Radiation properties of
  the accretion disk around a black hole in Einstein-Maxwell-scalar theory}},\
  }\href {https://doi.org/10.1088/1674-1137/ad137f} {\bibfield  {journal}
  {\bibinfo  {journal} {Chin. Phys. C}\ }\textbf {\bibinfo {volume} {48}},\
  \bibinfo {eid} {025101} (\bibinfo {year} {2024})},\ \Eprint
  {https://arxiv.org/abs/2401.03184} {arXiv:2401.03184 [gr-qc]} \BibitemShut
  {NoStop}%
\end{thebibliography}%
\end{document}